\documentclass[aps,pra,twocolumn,nofootinbib,superscriptaddress]{revtex4-2}

\usepackage[english]{babel}
\usepackage{amsmath}
\usepackage{amsthm}
\usepackage{amssymb}
\usepackage{booktabs}
\usepackage{color}
\usepackage{hyperref}
\hypersetup{
  colorlinks   = true,  
  urlcolor     = blue,  
  linkcolor    = blue,  
  citecolor    = red    
}
\usepackage{natbib}
\usepackage{pgfplots}
\usepackage{subfigure}
\usepackage{url}
\usepackage[normalem]{ulem}

\newcommand{\refEquation}[1]{{\textrm{Eq.~(\ref{#1})}}}

\usepackage{scalerel,stackengine}
\stackMath
\newcommand\reallywidehat[1]{%
\savestack{\tmpbox}{\stretchto{%
  \scaleto{%
    \scalerel*[\widthof{\ensuremath{#1}}]{\kern-.6pt\bigwedge\kern-.6pt}%
    {\rule[-\textheight/2]{1ex}{\textheight}}
  }{\textheight}%
}{0.5ex}}%
\stackon[1pt]{#1}{\tmpbox}%
}
\usepackage{comment}
\usepackage{float}
\usepackage{bm}

\begin{document}

\title{Three-body spin mixing in spin-1 Bose-Einstein condensates}

\author{P. M. A. \surname{Mestrom}}
\altaffiliation[Corresponding author: ]{p.m.a.mestrom@tue.nl}
\affiliation{Eindhoven University of Technology, P.~O.~Box 513, 5600 MB Eindhoven, The Netherlands}

\author{J.-L. \surname{Li}}
\affiliation{Eindhoven University of Technology, P.~O.~Box 513, 5600 MB Eindhoven, The Netherlands}

\author{V. E. \surname{Colussi}}
\affiliation{INO-CNR BEC Center and Dipartimento di Fisica, Universit\`{a} di Trento, 38123 Povo, Italy}

\author{T. \surname{Secker}}
\affiliation{Eindhoven University of Technology, P.~O.~Box 513, 5600 MB Eindhoven, The Netherlands}

\author{S. J. J. M. F. \surname{Kokkelmans}}
\affiliation{Eindhoven University of Technology, P.~O.~Box 513, 5600 MB Eindhoven, The Netherlands}

\date{\today}


\begin{abstract}
We study zero-energy collisions between three identical bosons with spin $f = 1$ interacting via pairwise potentials. We quantify the corresponding three-body scattering hypervolumes, which parametrize the effective three-body interaction strengths in a many-body description of spin-1 Bose-Einstein condensates. Our results demonstrate universal behavior of the scattering hypervolumes for strong $s$- and $p$-wave two-body interactions. At weak interactions we find that the real parts of the scattering hypervolumes are predominantly determined by hard-hyperspherelike collisions which we characterize by a simple formula. With this universal result we estimate that spin mixing via three-body collisions starts to dominate over two-body spin mixing at a typical particle density of $10^{17}~\mathrm{cm}^{-3}$ for ${}^{23}$Na and ${}^{41}$K spinor condensates. This density can be reduced by tuning the two-body interactions to an $s$- or $p$-wave dimer resonance or to a point where two-body spin mixing effectively vanishes. Another possibility to observe effects of three-body spin mixing involves the application of weak magnetic fields to cancel out the effective two-body interaction strength in the characteristic timescale describing the spin dynamics.
\end{abstract}

\maketitle

\section{Introduction}

Ultracold atoms provide a highly controllable platform to investigate quantum fluids with internal degrees of freedom. By using optical traps, one can create spinor Bose-Einstein condensates (BECs) in which the atoms are free to occupy different spin states. This spin degree of freedom gives rise to magnetic order, spin textures and nontrivial spin dynamics \cite{kawaguchi2012reviewSpinorBECs, stamperkurn2013reviewSpinorBoseGas}. The first spinor BEC consisting of ${}^{23}$Na atoms featured antiferromagnetic interactions \cite{stamperkurn1998spinorBEC,stenger1998spinorBEC}. Spinor condensates with ferromagnetic interactions have been created with ${}^{87}$Rb atoms \cite{chang2004spinorBEC87Rb} and recently also with ${}^{7}$Li atoms \cite{huh2020spinorBEC7Li}. The properties of spinor BECs depend crucially on spin-mixing collisions which change the spin state of the colliding atoms. Such collisions have been observed for several atomic species and spins \cite{stenger1998spinorBEC, chang2004spinorBEC87Rb, schmaljohann2004spinorBECRb87F2, kuwamoto2004spinorBECRb87F2, black2007spinorBECNa23, pasquiou2011spinorBECCr52, huh2020spinorBEC7Li, evrard2021spinorBECdynamicsNa23}.

The thermodynamical and non-equilibrium magnetic properties of dilute spinor condensates are predominantly determined by two-body spin-mixing collisions \cite{ho1998spinorBEC,ohmi1998spinorBEC,law1998spinorBEC,pethick2002bec,kawaguchi2012reviewSpinorBECs,stamperkurn2013reviewSpinorBoseGas}. Effects of three-particle collisions are usually expected to be weak although such effects have not been quantified. However, it is predicted that resonantly interacting spinor condensates can exhibit strong three-body effects with manifestations of Efimov physics \cite{colussi2014EfimovSpinor,colussi2016EfimovSpinor}. The spin degree of freedom gives rise to multiple families of Efimov trimer states \cite{bulgac1976spinEfimov,colussi2014EfimovSpinor,colussi2016EfimovSpinor}. Alternatively, comparatively strong three-body effects can be expected near a point where the effects of two-body collisions are suppressed. Such a point has been recently analyzed for a single-component BEC \cite{mestrom2020hypervolumeVdW, zwerger2019phasetransition}. For spinor condensates, the analysis is complicated by the additional scattering processes that arise from the three-body spin structure. These include spin-mixing collisions and can therefore impact the magnetic character of spinor BECs and their spin dynamics.

To quantify spin mixing via three-body collisions in spinor condensates, one needs to calculate the corresponding effective three-body interaction strengths which characterize the long-range behavior of the three-body scattering wave function. These strengths can be extracted from the elastic three-body transition amplitude as was recently demonstrated for spinless particles
\cite{braaten2002diluteBEC,mestrom2019hypervolumeSqW,mestrom2020hypervolumeVdW,mestrom2021hypervolumeBBX}. Via this approach the significance of three-body collisions in ultracold quantum gases was established for resonant $s$- \cite{braaten2002diluteBEC,mestrom2019hypervolumeSqW} and $p$-wave \cite{mestrom2021hypervolumeBBX} interactions and for weak interactions \cite{mestrom2019hypervolumeSqW,mestrom2020hypervolumeVdW} where such collisions are predicted to stabilize a single-component BEC against collapse \cite{mestrom2020hypervolumeVdW, bulgac2002droplets, zwerger2019phasetransition, hu2020excitationsDroplet, hu2021droplets3bodyTemperature}. In spinor condensates, multiple three-body spin channels are degenerate, resulting in effective three-body interaction strengths associated with each channel that is symmetric under particle exchange \cite{colussi2014EfimovSpinor,colussi2016EfimovSpinor}. 

In this paper, we quantify the effective three-body interaction strengths for spin-1 BECs and analyze when spin mixing via three-body collisions is important on the many-body level. We start our analysis in Section~\ref{sec:theory} by identifying the relevant spin channels for two- and three-body collisions at zero energy and defining the three-body scattering hypervolumes that parametrize the three-body effective interaction strengths for spin-1 BECs. In Section~\ref{sec:results_spin_1} we study the behavior of these scattering hypervolumes for resonant $s$- and $p$-wave interactions and for weak interactions. In Section~\ref{sec:spinor_condensates}, we input our findings into a many-body model and investigate the impact of three-body spin mixing on atomic spinor condensates both in and out of equilibrium.
Finally, we conclude in Section~\ref{sec:conclusion_spinor}.

\section{Scattering theory for identical spin-1 bosons}
\label{sec:theory}

In this section we investigate collisions between identical bosons with spin $f = 1$. First, we discuss the spin structure corresponding to one, two and three particles and identify the spin channels for scattering at zero collision energy. Next, we analyze the transition amplitudes for these two- and three-particle collisions. From these transition amplitudes, we define the scattering quantities that serve as inputs in the many-body theory of the spin-1 BEC developed later in Section~\ref{sec:spinor_condensates}. 



\subsection{Spin structure}


For identical particles with spin $f = 1$, the single-particle eigenstates $\lvert f, m_f \rangle$ are three-fold degenerate with $m_f = -1$, 0 or 1 being the corresponding magnetic quantum number. The eigenstates of the two-body spin Hamiltonian are given by 
\begin{equation}
\begin{aligned}
\lvert F_{\mathrm{2b}}, M_{F_{\mathrm{2b}}} \rangle 
= \sum_{m_{f_1}, m_{f_2}} &\langle f_1 m_{f_1} f_2 m_{f_2}| F_{\mathrm{2b}}, M_{F_{\mathrm{2b}}} \rangle  
\\
&\lvert f_1, m_{f_1} \rangle  \lvert f_2, m_{f_2}\rangle,
\end{aligned}
\end{equation}
where $\langle f_1 m_{f_1} f_2 m_{f_2}| F_{\mathrm{2b}}, M_{F_{\mathrm{2b}}} \rangle$ are the usual Clebsch-Gordan coefficients and $f_1$ and $f_2$ represent the spin of the two particles. $F_{\mathrm{2b}}$ and $M_{F_{\mathrm{2b}}} = m_{f_1} + m_{f_2}$ are the quantum numbers for the total two-body spin and its projection on the quantization axis, respectively. The states with $F_{\mathrm{2b}} = 0$ and 2 are symmetric under the permutation of the two spins, whereas the states $\lvert F_{\mathrm{2b}} = 1, M_{F_{\mathrm{2b}}} \rangle$ are antisymmetric. Therefore, two identical bosons with spin $f = 1$ can only collide at zero energy with $F_{\mathrm{2b}} = 0$ and 2. Such collisions play a crucial role in the phase diagram of spin-1 BECs \cite{kawaguchi2012reviewSpinorBECs, stamperkurn2013reviewSpinorBoseGas}.

To analyze the effects of three-body collisions on spinor condensates, we follow Refs.~\cite{colussi2014EfimovSpinor,colussi2016EfimovSpinor} and introduce the three-body spin states $\lvert F_{\mathrm{3b}}, M_{F_{\mathrm{3b}}} (F_{\beta \gamma}) \rangle_{\alpha}$ by
\begin{equation}
\begin{aligned}
\lvert F_{\mathrm{3b}}, M_{F_{\mathrm{3b}}} (F_{\beta \gamma}) \rangle_{\alpha} &\equiv
\lvert f_{\alpha}, f_{\beta}, f_{\gamma}, F_{\beta \gamma}, F_{\mathrm{3b}}, M_{F_{\mathrm{3b}}} \rangle_{\alpha}
\\
= \sum_{M_{F_{\beta \gamma}}, m_{f_{\alpha}}} &\langle F_{\beta \gamma} M_{F_{\beta \gamma}} f_{\alpha} m_{f_{\alpha}} | F_{\mathrm{3b}}, M_{F_{\mathrm{3b}}} \rangle  
\\
&\lvert F_{\beta \gamma}, M_{F_{\beta \gamma}} \rangle_{\alpha}  \lvert f_{\alpha} , m_{f_{\alpha}}\rangle_{\alpha},
\end{aligned}
\end{equation}
where $(\alpha,\beta,\gamma) = (1,2,3)$, $(3,1,2)$ or $(2,3,1)$ labels the particles, $f_{\alpha}$ is the spin of particle $\alpha$, $F_{\beta \gamma}$ is the $F_{\mathrm{2b}}$ corresponding to particles $\beta$ and $\gamma$, $F_{\mathrm{3b}}$ is the total three-body spin, and $M_{F_{\mathrm{3b}}} = m_{f_1} + m_{f_2} + m_{f_3}$ is its projection. The quantum numbers $F_{\mathrm{3b}}$ and $M_{F_{\mathrm{3b}}}$ are conserved in a three-body collision as we will see in Section~\ref{sec:theory_spinor_3body}. However, this is generally not true for $F_{\mathrm{2b}}$. 
Since we consider identical bosons, it is convenient to define the three-body spin states
\begin{equation}\label{eq:+_to_0_2_F3b=1}
\lvert 1, M_{F_{\mathrm{3b}}}[+] \rangle_{\alpha} \equiv \frac{\sqrt{5}}{3} \lvert 1, M_{F_{\mathrm{3b}}} (0) \rangle_{\alpha} + \frac{2}{3} \lvert 1, M_{F_{\mathrm{3b}}} (2) \rangle_{\alpha}
\end{equation}
and
\begin{equation}\label{eq:-_to_0_2_F3b=1}
\lvert 1, M_{F_{\mathrm{3b}}}[-] \rangle_{\alpha} \equiv \frac{2}{3} \lvert 1, M_{F_{\mathrm{3b}}} (0) \rangle_{\alpha} - \frac{\sqrt{5}}{3} \lvert 1, M_{F_{\mathrm{3b}}} (2) \rangle_{\alpha}.
\end{equation}
The states $\lvert 1, M_{F_{\mathrm{3b}}}[+] \rangle_{\alpha}$ and $\lvert 3, M_{F_{\mathrm{3b}}}(2) \rangle_{\alpha}$ are fully symmetric under permutations of any two spins \cite{a_noteSymmetricSpinStatesIndex}, whereas the state $\lvert 0, 0 (1) \rangle_{\alpha}$ is antisymmetric. All other three-body spin states including $\lvert 1, M_{F_{\mathrm{3b}}}[-] \rangle_{\alpha}$ are not fully symmetric or antisymmetric. At zero energy, three identical spin-1 bosons can thus only collide beginning from incoming spin channels $\lvert 1, M_{F_{\mathrm{3b}}}[+] \rangle_{\alpha}$ and $\lvert 3, M_{F_{\mathrm{3b}}}(2) \rangle_{\alpha}$ which are three- and sevenfold degenerate, respectively. During such a collision, other spin states can be involved as well. In Section~\ref{sec:theory_spinor_3body} we detail which spin states are coupled.

\subsection{Two-body transition amplitude}
\label{sec:theory_spinor_2body}

Before discussing three-body scattering theory, we briefly comment on the considered pairwise interaction potentials and the corresponding two-body transition amplitude. We consider spin-1 particles that interact in pairs via an interaction operator $V$ that is spherically symmetric and conserves $F_{\mathrm{2b}}$ and $M_{F_{\mathrm{2b}}}$, i.e.,
\begin{equation}\label{eq:V_vs_VF2b}
\begin{aligned}
V =& \sum_{F_{\mathrm{2b}},M_{F_{\mathrm{2b}}}} \int \,d\mathbf{p} \,d\mathbf{p}' \,\lvert \mathbf{p}, F_{\mathrm{2b}},M_{F_{\mathrm{2b}}} \rangle \langle \mathbf{p} | V_{F_{\mathrm{2b}}} | \mathbf{p}'\rangle
\\
& \langle \mathbf{p}', F_{\mathrm{2b}},M_{F_{\mathrm{2b}}} \rvert.
\end{aligned}
\end{equation}
Here $V_{F_{\mathrm{2b}}}$ is the interaction operator between two particles with total spin $F_{\mathrm{2b}}$ and $\mathbf{p}$ and $\mathbf{p}'$ represent relative momenta between the two particles. Throughout this paper we normalize plane wave states according to $\langle \mathbf{p}'| \mathbf{p} \rangle = \delta(\mathbf{p}' - \mathbf{p})$. 

The transition operator $t_{F_{\mathrm{2b}}}\left(z_{\mathrm{2b}}\right)$ describes two-body scattering processes at energy $z_{\mathrm{2b}}$. It is defined via the Lippmann-Schwinger equation
\begin{equation}\label{eq:LS_tF2b}
t_{F_{\mathrm{2b}}}\left(z_{\mathrm{2b}}\right) = V_{F_{\mathrm{2b}}} + V_{F_{\mathrm{2b}}} G_0^{(\mathrm{2b})}\left(z_{\mathrm{2b}}\right) t_{F_{\mathrm{2b}}}\left(z_{\mathrm{2b}}\right),
\end{equation}
where $G_0^{(\mathrm{2b})}\left(z_{\mathrm{2b}}\right) = \left(z_{\mathrm{2b}}-H_0^{(\mathrm{2b})}\right)^{-1}$ and $H_0^{(\mathrm{2b})}$ is the two-body kinetic energy operator in the center-of-mass frame. For zero-energy collisions with spin $F_{\mathrm{2b}}$, the scattering cross section is completely determined by the scattering length $a_{F_{\mathrm{2b}}}$ \cite{taylor1972scattering} which is defined by the two-body transition amplitude $\langle \mathbf{p} | t_{F_{\mathrm{2b}}}\left(z_{\mathrm{2b}}\right) | \mathbf{p}' \rangle$ via
\begin{equation}
a_{F_{\mathrm{2b}}} = 2 \pi^2 m \hbar \langle \mathbf{0} | t_{F_{\mathrm{2b}}}(0) | \mathbf{0} \rangle.
\end{equation}
Here $m$ is the mass of a particle and should not be confused with the quantum number $m_f$.

\subsection{Three-body transition amplitude}
\label{sec:theory_spinor_3body}

The Alt-Grassberger-Sandhas (AGS) equations \cite{alt1967ags},
\begin{equation}\label{eq:AGS_U_alpha0_spinor}
\begin{aligned}
U_{\alpha 0}(z) &= (1-\delta_{\alpha 0}) G_0^{-1}(z) + \sum_{\substack{\beta = 1 \\ \beta \neq \alpha}}^{3} T_{\beta}(z) G_0(z) U_{\beta 0}(z)
\\
&\text{ for } \alpha = 0, 1, 2, 3,
\end{aligned}
\end{equation}
define the transition operators $U_{\alpha 0}(z)$ for scattering of three free particles at energy $z$. The outgoing state is labeled by $\alpha$ and consists of three free particles ($\alpha = 0$) or a free particle $\alpha$ and a $\beta \gamma$ dimer in which case $(\alpha,\beta,\gamma) = (1,2,3)$, $(3,1,2)$ or $(2,3,1)$.
The Green's function $G_0(z) = \left(z- H_0\right)^{-1}$ contains the three-body kinetic energy operator $H_0$ in the center-of-mass frame. The transition operator $T_{\alpha}(z)$ with $\alpha = 1$, 2 or 3 is defined via $T_{\alpha}(z) = V_{\beta \gamma} + V_{\beta \gamma} G_0(z) T_{\alpha}(z)$ where the pairwise interaction $V_{\beta \gamma}$ between particles $\beta$ and $\gamma$ is given in Eq.~\eqref{eq:V_vs_VF2b}.
It is connected to $t_{F_{\mathrm{2b}}}$ via
\begin{equation}
\begin{aligned}
T_{\alpha}&(z) = \sum_{F_{\mathrm{3b}}, M_{F_{\mathrm{3b}}},F_{\beta \gamma}} 
\int \,d\mathbf{q}_{\alpha} \,d\mathbf{p}_{\alpha} \,d\mathbf{p}_{\alpha}'
\\
&\lvert \mathbf{p}_{\alpha}, \mathbf{q}_{\alpha}, F_{\mathrm{3b}}, M_{F_{\mathrm{3b}}} (F_{\beta \gamma}) \rangle_{\alpha} \,
\langle \mathbf{p}_{\alpha} | t_{F_{\beta \gamma}}\left(z-\frac{3 q_{\alpha}^2}{4 m}\right) | \mathbf{p}_{\alpha}'\rangle
\\
&\,{}_{\alpha}\langle \mathbf{p}_{\alpha}', \mathbf{q}_{\alpha}, F_{\mathrm{3b}}, M_{F_{\mathrm{3b}}} (F_{\beta \gamma}) \rvert,
\end{aligned}
\end{equation}
where the plane wave states $\lvert \mathbf{p}_{\alpha}, \mathbf{q}_{\alpha} \rangle_{\alpha}$ describe the relative motion of the three-body system. The relative momenta $\mathbf{p}_{\alpha}$ and $\mathbf{q}_{\alpha}$ are the Jacobi momenta and are defined by the lab momenta $\mathbf{P}_{\alpha}$ of particles $\alpha = 1$, 2 and 3 via $\mathbf{p}_{\alpha} = (\mathbf{P}_{\beta} - \mathbf{P}_{\gamma})/2$ and $\mathbf{q}_{\alpha}~=~(2/3) \left[ \mathbf{P}_{\alpha} - \left(\mathbf{P}_{\beta} + \mathbf{P}_{\gamma}\right)/2\right]$.

It is useful to define the operator $\breve{U}_{\alpha 0}(z) \equiv T_{\alpha}(z) G_0(z) U_{\alpha 0}(z) (1+P)$ for $\alpha = 1,2,3$ where $P$ is the sum of the cyclic and anticyclic permutation operators. 
For identical particles, we derive from \refEquation{eq:AGS_U_alpha0_spinor} that $\breve{U}_{\alpha 0}(z)$ is determined by
\begin{equation}\label{eq:Ubreve_BBB_FFF}
\breve{U}_{\alpha 0}(z) = T_{\alpha}(z)(1+P) + T_{\alpha}(z) G_0(z) P \breve{U}_{\alpha 0}(z).
\end{equation}
The zero-energy three-body transition amplitude can thus be expressed by
\begin{equation}\label{eq:U00_vs_Ubreve_matrix_Spin}
\begin{aligned}
\langle \mathbf{p}, \mathbf{q}, \Sigma_{\mathrm{3b}} | &U_{00}(0) | \mathbf{0}, \mathbf{0}, \Sigma_{\mathrm{3b,in}} \rangle 
\\
&= \frac{1}{3} \sum_{\alpha = 1}^{3} {}_{\alpha}\langle \mathbf{p}_{\alpha}, \mathbf{q}_{\alpha}, \Sigma_{\mathrm{3b}} | \breve{U}_{\alpha 0}(0) | \mathbf{0}, \mathbf{0}, \Sigma_{\mathrm{3b,in}} \rangle,
\end{aligned}
\end{equation}
where we take $z \to 0$ from the upper half of the complex energy plane and define $\mathbf{p} \equiv \mathbf{p}_1$ and $\mathbf{q} \equiv \mathbf{q}_1$. Since the spatial part $\lvert \mathbf{0}, \mathbf{0} \rangle$ of the incoming state is fully symmetric under any permutation of the particles, the incoming spin channel $\lvert \Sigma_{\mathrm{3b,in}} \rangle$ must be one of the symmetric states $\lvert 1, M_{F_{\mathrm{3b}}}[+] \rangle$ and $\lvert 3, M_{F_{\mathrm{3b}}}(2) \rangle$ for identical bosons.
The operators $G_0(z)$ and $T_{\alpha}(z)$ conserve $F_{\mathrm{3b}}$, $M_{F_{\mathrm{3b}}}$ and $F_{\beta \gamma}$, whereas $P$ conserves $F_{\mathrm{3b}}$ and $M_{F_{\mathrm{3b}}}$. However, $P$ also conserves $F_{\mathrm{2b}} = 2$ for $F_{\mathrm{3b}} = 3$, so that the outgoing spin state $\lvert \Sigma_{\mathrm{3b}} \rangle$ equals $\lvert \Sigma_{\mathrm{3b,in}} \rangle$ for $\lvert \Sigma_{\mathrm{3b,in}} \rangle = \lvert 3, M_{F_{\mathrm{3b}}} (2)\rangle$ and the corresponding three-body equations map onto those for spinless bosons.
Therefore, the three-body scattering wave function of three identical spin-1 bosons in the spin state $\lvert 3, M_{F_{\mathrm{3b}}} (2)\rangle$ is identical to the one for identical spinless bosons interacting via the pairwise interaction potential $V_{F_{\mathrm{2b}} = 2}$. The latter has been studied at zero collision energy in Refs.~\cite{mestrom2019hypervolumeSqW, mestrom2020hypervolumeVdW} for various finite-range potentials including van der Waals potentials.

For identical spin-1 bosons with $F_{\mathrm{3b}} = 1$, the situation is different because $V$ is not diagonal in the spin states $\lvert 1, M_{F_{\mathrm{3b}}} [\pm]\rangle$. Since the permutation operator $P$ couples the spin states $\lvert 1, M_{F_{\mathrm{3b}}} [-]\rangle$ and $\lvert 1, M_{F_{\mathrm{3b}}} (1)\rangle$, the three-body scattering wave function depends on $V_{F_{\mathrm{2b}} = 0}$, $V_{F_{\mathrm{2b}} = 1}$ and $V_{F_{\mathrm{2b}} = 2}$. 

The elastic three-body transition amplitude corresponding to scattering at zero energy in the spin channels $\lvert \Sigma_{\mathrm{3b}} \rangle = \lvert 1, M_{F_{\mathrm{3b}}}[+] \rangle$ and $\lvert 3, M_{F_{\mathrm{3b}}}(2) \rangle$ can be written as
\begin{equation}\label{eq:U00_momentum_Phi3b}
\begin{aligned}
&\langle \mathbf{p}, \mathbf{q}, \Sigma_{\mathrm{3b}} | U_{0 0}(0) | \mathbf{0}, \mathbf{0}, \Sigma_{\mathrm{3b}} \rangle = \sum_{\alpha = 1}^{3}\Bigg\{\delta(\mathbf{q}_{\alpha}) 
\\
&\times \sum_{F_{\beta \gamma}} \left|Z_{F_{\beta \gamma},\Sigma_{\mathrm{3b}}}\right|^2 \langle \mathbf{p}_{\alpha} | t_{F_{\beta \gamma}}(0) | \mathbf{0} \rangle 
+ \frac{A_{F_{\mathrm{3b}}}}{q_{\alpha}^2}
+ \frac{B_{F_{\mathrm{3b}}}}{q_{\alpha}}
\\
&+ C_{F_{\mathrm{3b}}} \, \text{ln}\bigg( \frac{q_{\alpha} \rho_{F_{\mathrm{3b}}}}{\hbar} \bigg)
+ \frac{1}{(2 \pi)^6} \mathcal{U}^{(\alpha)}_{\Sigma_{\mathrm{3b}}}(\mathbf{p}_{\alpha},\mathbf{q}_{\alpha}) \Bigg\},
\end{aligned}
\end{equation}
where $\rho_{F_{\mathrm{3b}}}$ is an arbitrary, positive length scale and
\begin{equation}\label{eq:def_Z_F2b_Phi3b}
Z_{F_{\beta \gamma},\Sigma_{\mathrm{3b}}} = {}_{\alpha}\langle F_{\mathrm{3b}}, M_{F_{\mathrm{3b}}} (F_{\beta \gamma}) | \Sigma_{\mathrm{3b}} \rangle.
\end{equation}
Furthermore, $A_{F_{\mathrm{3b}}}$, $B_{F_{\mathrm{3b}}}$ and $C_{F_{\mathrm{3b}}}$ are real coefficients depending on the scattering lengths $a_{F_{\mathrm{2b}}}$ and are given in Appendix~\ref{sec:U00_amplitude_spinor}. The functions $\mathcal{U}^{(\alpha)}_{\Sigma_{\mathrm{3b}}}(\mathbf{0},\mathbf{q}_{\alpha})$ are nonsingular in $q_{\alpha} = 0$ and do not depend on $M_{F_{\mathrm{3b}}}$ because the Hamiltonian is invariant with respect to rotations of the complete three-body system. We can thus define
\begin{equation}\label{eq:U0_def_Phi3b}
\begin{aligned}
\mathcal{U}_0^{(F_{\mathrm{3b}})} &= \sum_{\alpha = 1}^{3} \lim_{q_{\alpha} \to 0} \mathcal{U}^{(\alpha)}_{\Sigma_{\mathrm{3b}}}(\mathbf{0},\mathbf{q}_{\alpha})
\\
&= 3 \, \mathcal{U}^{(\alpha)}_{\Sigma_{\mathrm{3b}}}(\mathbf{0},\mathbf{0}),
\end{aligned}
\end{equation}
where in the final line $\alpha = 1$, 2 or 3 is arbitrary for identical particles. From $\mathcal{U}_0^{(F_{\mathrm{3b}})}$ we define the three-body scattering hypervolume $D_{F_{\mathrm{3b}}}$ by generalizing a previous definition for spinless bosons \cite{tan2008hypervolume,mestrom2019hypervolumeSqW}. This results in
\begin{equation}\label{eq:def_D_F3b_spinor}
\begin{aligned}
D_{F_{\mathrm{3b}}} &= m\hbar^4 \mathcal{U}_0^{(F_{\mathrm{3b}})} + 12 \pi^4 m \hbar^3 \sum_{\alpha = 1}^{3} \sum_{F_{\beta \gamma}} \left(Z_{F_{\beta \gamma},\Sigma_{\mathrm{3b}}}\right)^* 
\\
&\times \frac{\partial^2 \langle \mathbf{p} | t_{F_{\beta \gamma}}(0) | \mathbf{0} \rangle}{\partial p^2}\Bigg|_{p=0}
 \sum_{F_{\beta \gamma}'} W_{F_{\beta \gamma},F_{\beta \gamma}'}^{(F_{\mathrm{3b}})} a_{F_{\beta \gamma}'}Z_{F_{\beta \gamma}',\Sigma_{\mathrm{3b}}} 
\end{aligned}
\end{equation}
where we have defined
\begin{equation}\label{eq:def_W_F2b_F2bprime^(F3b)}
W_{F_{\beta \gamma},F_{\beta \gamma}'}^{(F_{\mathrm{3b}})} = 2 \, {}_{\alpha}\langle F_{\mathrm{3b}}, M_{F_{\mathrm{3b}}} (F_{\beta \gamma}) |P_+^{\mathrm{s}} | F_{\mathrm{3b}}, M_{F_{\mathrm{3b}}} (F_{\beta \gamma}') \rangle_{\alpha}
\end{equation}
and $P_+^{\mathrm{s}}$ is the cyclic permutation operator acting on spin space. In Appendix~\ref{sec:U00_amplitude_spinor} we specify $W_{F_{\beta \gamma},F_{\beta \gamma}'}^{(F_{\mathrm{3b}})}$ for $F_{\mathrm{3b}} = 1$ and 3. We note that the definition of $\mathcal{U}_0^{(F_{\mathrm{3b}})}$ and $D_{F_{\mathrm{3b}}}$ is only fixed when $\rho_{F_{\mathrm{3b}}}$ is specified. For $F_{\mathrm{3b}} = 3$ we find that $D_3$ is identical to the three-body scattering hypervolume $D$ defined in Refs.~\cite{tan2008hypervolume,mestrom2019hypervolumeSqW} for spinless bosons interacting via pairwise potentials $V_{F_{\mathrm{2b}} = 2}$ when $\rho_{3} = |a_2|$. Furthermore, $D_1 = D_3$ for $\rho_1 = \rho_3$ when $V_{F_{\mathrm{2b}} = 0} = V_{F_{\mathrm{2b}} = 2}$ in which case $T_{\alpha}(z)$ is diagonal in the spin states $\lvert 1, M_{F_{\mathrm{3b}}} [\pm]\rangle$. We also note that $\mathrm{Im}\left(D_{F_{\mathrm{3b}}}\right)$ is not affected by the choice for $\rho_{F_{\mathrm{3b}}}$.

To calculate $D_1$ and $D_3$, we solve Eq.~\eqref{eq:Ubreve_BBB_FFF} in momentum space using the same method as in our previous works for spinless particles \cite{mestrom2019hypervolumeSqW,mestrom2020hypervolumeVdW,mestrom2021hypervolumeBBX}. The corresponding integral equations are presented in Appendix~\ref{sec:integral_eq_spinor}. They are connected to the integral equations for spinless particles in Appendix~\ref{sec:spinor_connection_BBX}. In Section~\ref{sec:results_spin_1} we analyze the behavior of $D_1$ and $D_3$ for various interaction potentials $V_{F_{\mathrm{2b}}}$.

Finally, we note that the imaginary parts of $D_1$ and $D_3$ are proportional to the three-body recombination rate at zero energy in a similar way as for spinless bosons \cite{tan2017hypervolume}. This follows directly from the optical theorem for three-body scattering \cite{schmid1974threebody}. On the other hand, the real parts of $D_1$ and $D_3$ are determined by elastic scattering processes. They play a similar role as the scattering lengths $a_0$ and $a_2$ in the description of spin-1 BECs. This connection to the properties of spinor condensates is discussed in Section~\ref{sec:MF_spinor_BEC}.

\section{Three-body scattering hypervolumes}
\label{sec:results_spin_1}

The three-body scattering hypervolume $D_3$ can be obtained directly from the three-body scattering hypervolume for identical spinless bosons which has been carefully analyzed in Refs.~\cite{tan2008hypervolume,tan2017hypervolume,mestrom2019hypervolumeSqW,mestrom2020hypervolumeVdW} for finite-range potentials. However, $D_1$ has not been quantified before. Previous studies \cite{colussi2014EfimovSpinor,colussi2016EfimovSpinor} determined the qualitative behavior of $D_1$ for resonant $s$-wave interactions, i.e., $a_0 \to \pm \infty$ and $a_2 \to \pm \infty$. In this section we quantify $D_1$ for resonant $s$- and $p$-wave interactions and for weak interactions in the regime that is relevant for several alkali-metal atoms. 
Although resonant interactions do not appear naturally in atomic systems in the absence of an external magnetic field, we will see in Sections~\ref{ssec:results_strong_swave} and \ref{ssec:results_strong_pwave} that they cause $D_1$ to diverge and thereby can lead to strong effects of three-body spin mixing on spinor condensates. Our analysis of $D_1$ for weak pairwise interactions in Section~\ref{ssec:results_weak_interactions} corresponds to the natural situation in ultracold alkali-metal atoms at zero magnetic field. We apply the results of this analysis to atomic spin-1 BECs in Section~\ref{sec:spinor_condensates}.

\subsection{Strong $s$-wave interactions}
\label{ssec:results_strong_swave}


Identical bosons with spin $f = 1$ can form an $s$-wave dimer state with $F_{\mathrm{2b}} = 0$ or 2. At resonance, the corresponding two-body scattering length $a_{F_{\mathrm{2b}}}$ diverges. In the limits $|a_0| \to \infty$ and $|a_2| \to \infty$, the three-body scattering hypervolume $D_1$ diverges as $a_0^4$ and $a_2^4$, respectively \cite{colussi2014EfimovSpinor}. To quantify these limits, we take contact interactions for $V_{F_{\mathrm{2b}}}$ with a momentum cut-off $\Lambda$, i.e.,
\begin{equation} \label{eq:V_sepCut_F2b}
V_{F_{\mathrm{2b}}} = - \zeta_{F_{\mathrm{2b}}} \lvert g \rangle \langle g \rvert,
\end{equation}
where
\begin{equation}
\langle \mathbf{p} | g \rangle = 
\begin{cases}1,&\mbox{$0\leq p \leq \Lambda$},\\0,&\mbox{$p> \Lambda$}.\end{cases}
\end{equation}
We set $\zeta_{1} = 0$ and tune $\zeta_{0}$ and $\zeta_{2}$ to vary the scattering lengths $a_0$ and $a_2$, respectively. With this pairwise interaction, we numerically find the universal limits
\begin{equation}\label{eq:ReD1_a0_inf}
\begin{aligned}
\mathrm{Re}\left(D_1\right)/a_0^4 \underset{a_0 \to \pm\infty}{=}& \frac{640 \pi^2}{81} \,\text{ln}(|a_{0}|/\rho_1) \left(\frac{2}{3} - \frac{3\sqrt{3}}{2 \pi}\right)
\\
&+ 5.485(5)
\end{aligned}
\end{equation}
and
\begin{equation}\label{eq:ReD1_a2_inf}
\begin{aligned}
\mathrm{Re}\left(D_1\right)/a_2^4 \underset{a_2 \to \pm\infty}{=}& \frac{128 \pi^2}{81} \,\text{ln}(|a_{2}|/\rho_1) \left(\frac{1}{3} - \frac{3\sqrt{3}}{2 \pi}\right)
\\
&- 24.77(1).
\end{aligned}
\end{equation}
For the imaginary part we find 
\begin{equation}\label{eq:ImD1_a0_inf}
\begin{aligned}
\mathrm{Im}\left(D_1\right)/a_0^4 &\underset{a_0 \to +\infty}{=} - 39.28(1)
\end{aligned}
\end{equation}
and
\begin{equation}\label{eq:ImD1_a2_inf}
\begin{aligned}
\mathrm{Im}\left(D_1\right)/a_2^4 &\underset{a_2 \to +\infty}{=} - 24.19(1),
\end{aligned}
\end{equation}
which set the three-body recombination rates into the shallow $s$-wave dimer state with $F_{\mathrm{2b}} = 0$ and 2, respectively. 

\begin{figure}[btp]
	\begin{subfigure}
	\centering
	\includegraphics[width=3.4in]{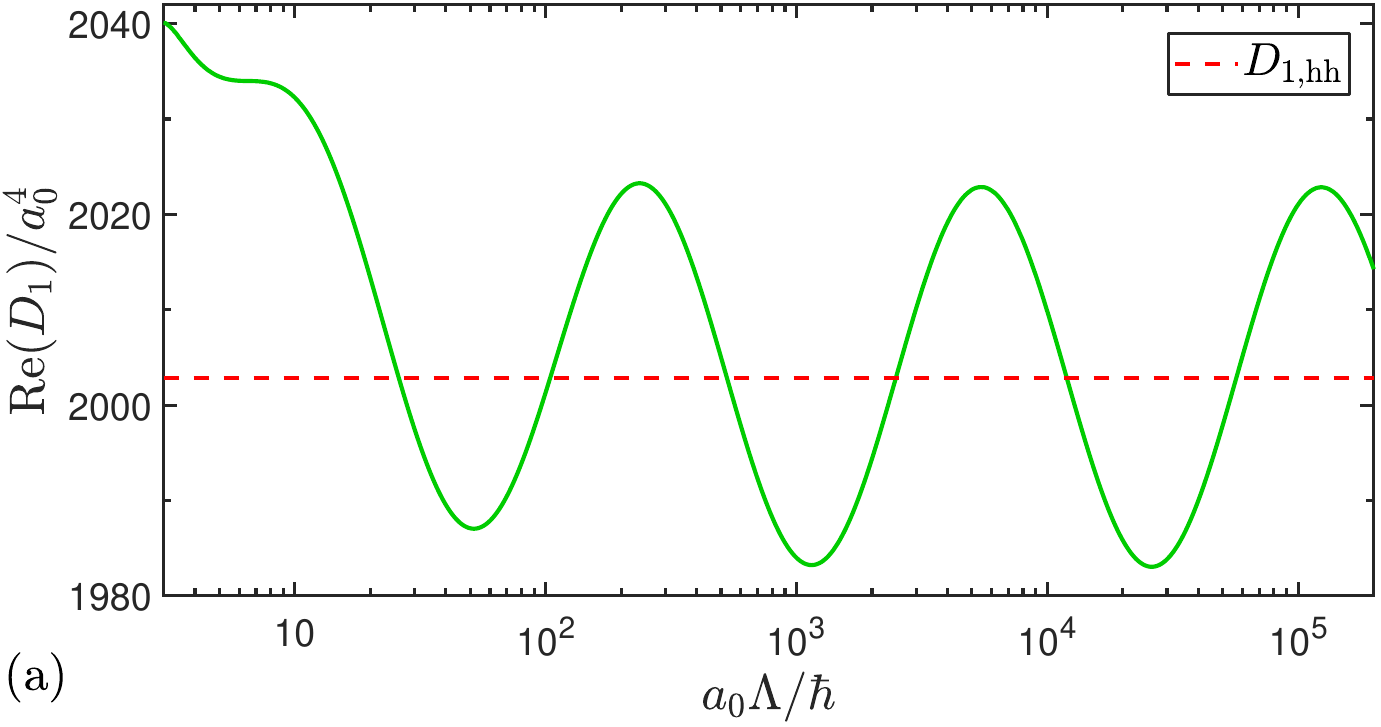}
	\end{subfigure}
\quad
	\begin{subfigure}
	\centering
	\includegraphics[width=3.4in]{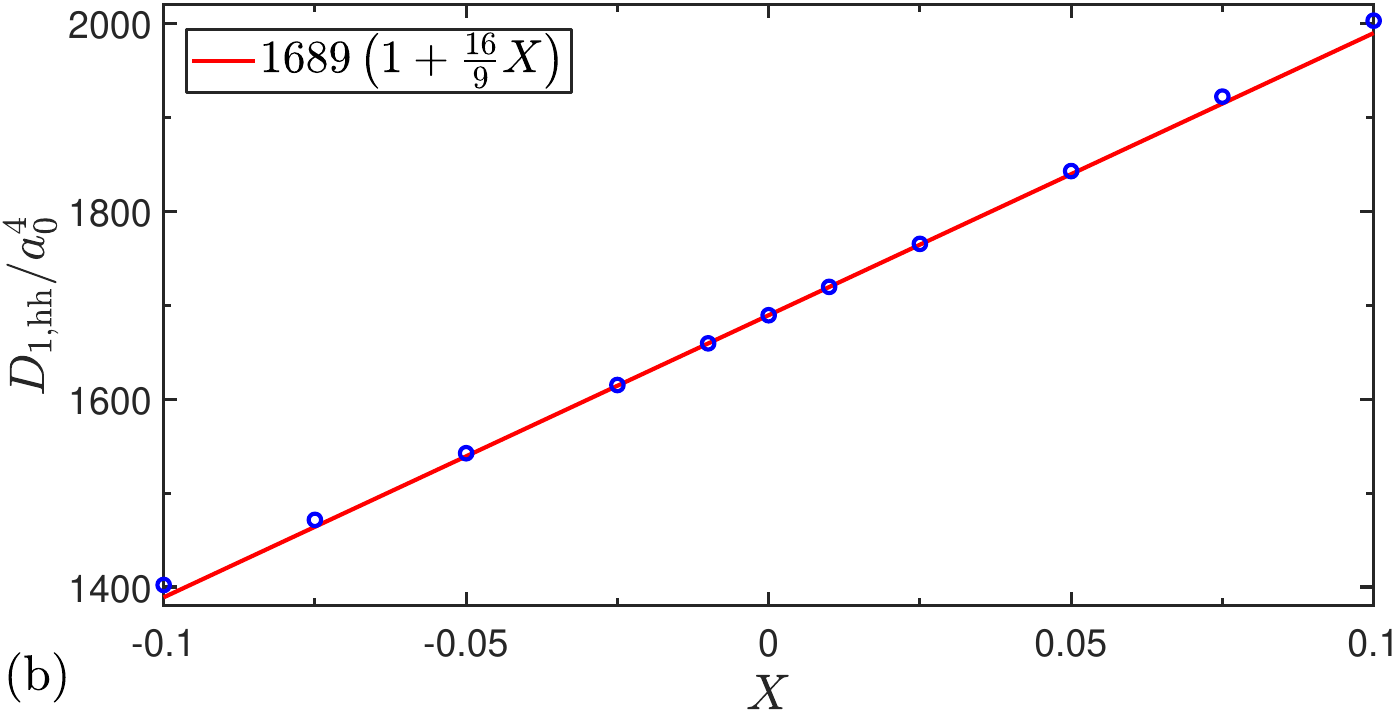}
	\end{subfigure}
    \caption{(a) $\mathrm{Re}\left(D_1\right)$ corresponding to the interaction potential in Eq.~\eqref{eq:V_sepCut_F2b} for $V_{F_{\mathrm{2b}} = 1} = 0$ and $a_2/a_0 = 1.1$. We fix the definition of $D_{1}$ by setting $\rho_1 = \lvert \tilde{a}\rvert$. The red dashed line indicates the offset $D_{1,\mathrm{hh}}$ around which $D_1$ oscillates for large $a_0 \Lambda/\hbar$ due to the Efimov effect. The amplitude of the oscillation is only 1\% of $D_{1,\mathrm{hh}}$. (b) Values of $D_{1,\mathrm{hh}}$ with $\rho_1 = \lvert \tilde{a}\rvert$ as a function of $X = (a_2-a_0)/a_0$ near $X = 0$. The red line shows the first-order Taylor approximation around $X = 0$ as given by \refEquation{eq:Dh_hh_F3b_1_Taylor_X}.}
    \label{fig:Dhh_F3b_1_vs_X}
\end{figure}

Recent studies \cite{colussi2014EfimovSpinor,colussi2016EfimovSpinor} demonstrated that the Efimov effect \cite{efimov1970energy,efimov1971weakly,bulgac1976spinEfimov} only occurs for three identical spin-1 bosons with $F_{\mathrm{3b}} = 1$ when $a_0$ and $a_2$ diverge simultaneously. For the specific case that $a_0 = a_2$ diverges, $D_1$ follows the universal relations for resonant $s$-wave interactions presented in Ref.~\cite{mestrom2019hypervolumeSqW} for the three-body scattering hypervolume of spinless bosons. This does not mean that $D_1 = D_3$ since the universal formulas depend on the short-range details of the interaction potentials via a three-body parameter that fixes the Efimov spectrum and an inelasticity parameter that determines the decay rate to deeply bound dimer states \cite{mestrom2019hypervolumeSqW}. However, the contribution to $D_1$ that comes from hard-hyperspherelike collisions \cite{dincao2018review,mestrom2020hypervolumeVdW} is not influenced by short-range details and depends only on $a_0$ and $a_2$ for resonant $s$-wave interactions. We call this contribution $D_{1,\mathrm{hh}}$. Its value is determined solely by scattering pathways in which the particles reflect off a barrier in the three-body effective potentials for the hyperradial motion \cite{dincao2018review}, hence the name hard-hyperspherelike collisions. From the analysis in Refs.~\cite{braaten2002diluteBEC,mestrom2019hypervolumeSqW, mestrom2020hypervolumeVdW}, we thus find that $D_{1,\mathrm{hh}} = D_{3,\mathrm{hh}} = 1689 a_2^4$ for $a_0 = a_2$ and $\rho_1 = \rho_3 = |a_2|$. This follows immediately from the fact that $D_1 = D_3$ for $V_{F_{\mathrm{2b}} = 0} = V_{F_{\mathrm{2b}} = 2}$ in which case the condition $a_0 = a_2$ is automatically fulfilled.
We can even go a step further and consider $a_2 = (1+ X) a_0$ for small $X$. By defining 
\begin{equation}\label{eq:atilde_def}
\tilde{a} = \frac{5}{9} a_0 + \frac{4}{9} a_2
\end{equation}
and taking $\rho_1 = |\tilde{a}|$, we find that
\begin{equation}\label{eq:Dh_hh_F3b_1_Taylor_X}
\begin{aligned}
D_{1,\mathrm{hh}} &= 1689 \, \tilde{a}^4 + O(X^2)
\\
&= 1689 \left(1 + \frac{16}{9} X \right) a_0^4 + O(X^2)
\\
&= 1689 \left(1 - \frac{20}{9} X \right) a_2^4 + O(X^2).
\end{aligned}
\end{equation}
A derivation can be found in Appendix~\ref{sec:special_cases}. Figure~\ref{fig:Dhh_F3b_1_vs_X} confirms the first-order Taylor approximation in \refEquation{eq:Dh_hh_F3b_1_Taylor_X}. This result can even be used to describe the behavior of $\mathrm{Re}\left(D_1\right)$ for weak interactions as we demonstrate in Section~\ref{ssec:results_weak_interactions}.

\begin{figure}[btp]
	\begin{subfigure}
	\centering
	\includegraphics[width=3.4in]{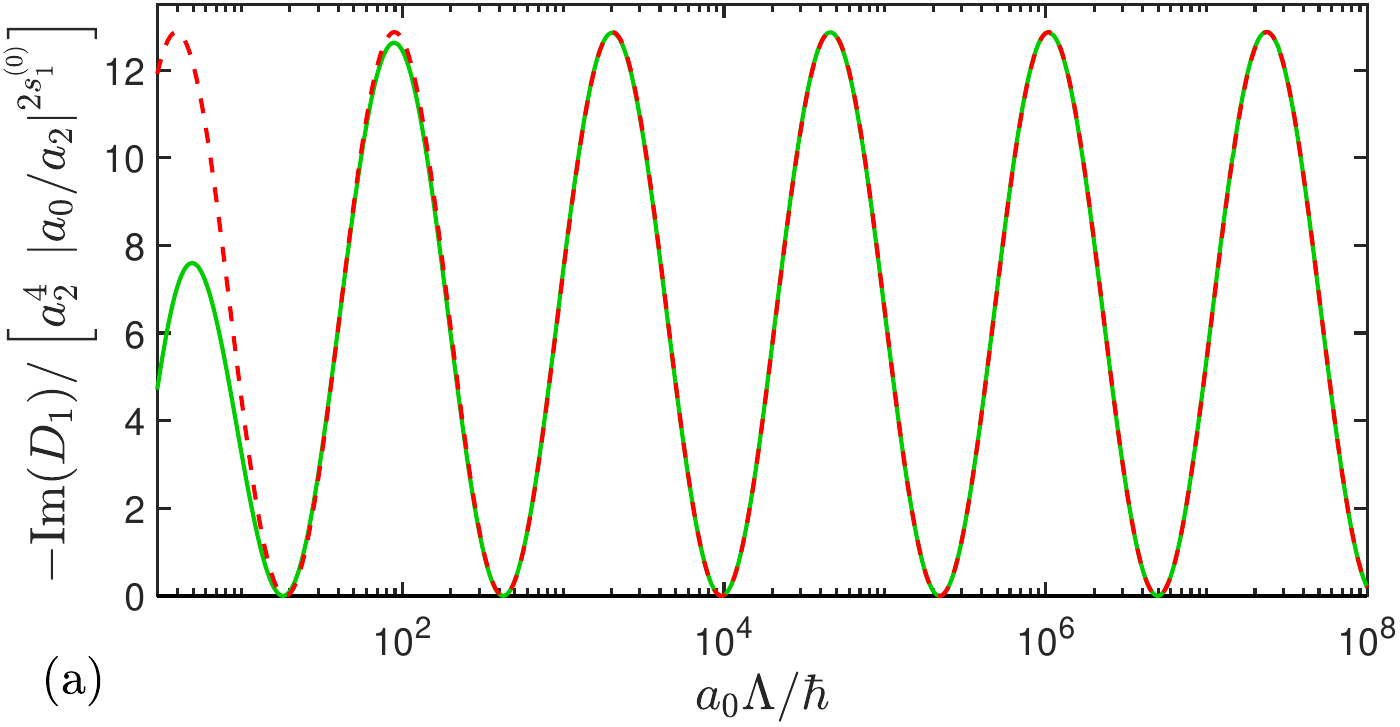}
	\end{subfigure}
\quad
	\begin{subfigure}
	\centering
	\includegraphics[width=3.4in]{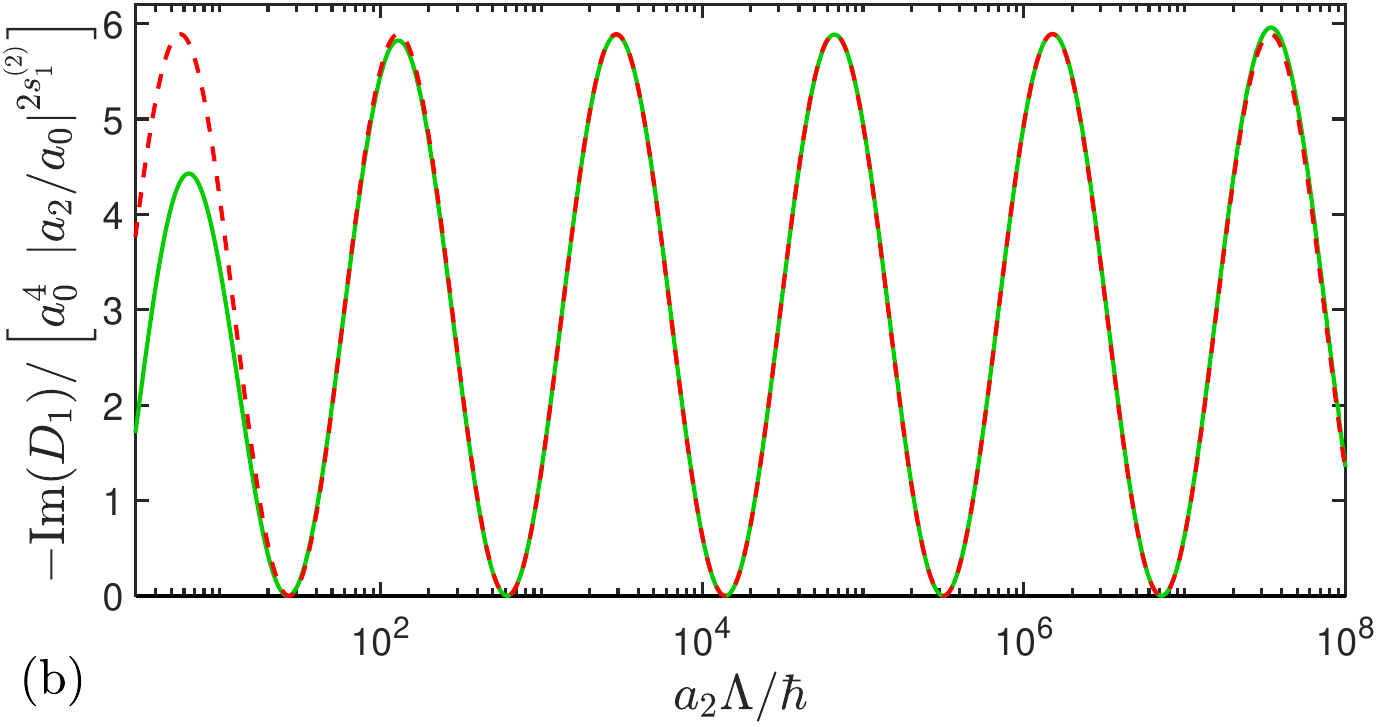}
	\end{subfigure}
    \caption{$\mathrm{Im}\left(D_1\right)$ corresponding to the interaction potential in Eq.~\eqref{eq:V_sepCut_F2b} for $V_{F_{\mathrm{2b}} = 1} = 0$. (a) We fix $a_2 \Lambda/\hbar = -10^{12}$ and vary $a_0$. The red dashed curve corresponds to Eq.~\eqref{eq:Im_D1_vs_a0_Large_neg_a2} with $a_{0,+} \Lambda/\hbar = 2.191\cdot 10^5$ and $C^{(0)} = 12.87$. (b) We fix $a_0 \Lambda/\hbar = -10^{12}$ and vary $a_2$.  The red dashed curve corresponds to Eq.~\eqref{eq:Im_D1_vs_a2_Large_neg_a0} with $a_{2,+} \Lambda/\hbar = 3.163\cdot 10^5$ and $C^{(2)} = 5.89$.   
    }
    \label{fig:Im_D1_large_a0_and_a2}
\end{figure}

For large values of $|X|$ or $X \approx -1$, we enter the regime that $|a_0|\ll |a_2|$ or $|a_2|\ll |a_0|$. The behavior of $D_1$ in the corresponding strongly interacting regime was studied in Ref.~\cite{colussi2014EfimovSpinor}. In Fig.~\ref{fig:Im_D1_large_a0_and_a2} we confirm several predictions of Ref.~\cite{colussi2014EfimovSpinor}, namely
\begin{equation}\label{eq:Im_D1_vs_a0_Large_neg_a2}
\mathrm{Im}\left(D_1\right)/a_2^4 = - C^{(0)} \sin^2\left[s_0\,\ln(a_0/a_{0,+})\right] \left|a_0/a_2\right|^{2 s_1^{(0)}},
\end{equation}
for $a_0 \ll |a_2|$ with $a_0>0$ and $a_2<0$ and
\begin{equation}\label{eq:Im_D1_vs_a2_Large_neg_a0}
\mathrm{Im}\left(D_1\right)/a_0^4 = - C^{(2)} \sin^2\left[s_0\,\ln(a_2/a_{2,+})\right] \left|a_2/a_0\right|^{2 s_1^{(2)}}
\end{equation}
for $a_2 \ll |a_0|$ with $a_0<0$ and $a_2>0$. Here $s_0 \approx 1.00624$, $s_1^{(0)} \approx 0.74289$ and $s_1^{(2)} \approx 0.40970$ \cite{colussi2014EfimovSpinor}. The three-body parameters $a_{0,+}$ and $a_{2,+}$ locate the minima of $-\mathrm{Im}\left(D_1\right)$ that originate from interfering pathways for recombination into the shallow dimer state with $F_{\mathrm{2b}} = 0$ and $F_{\mathrm{2b}} = 2$, respectively. The coefficients $C^{(0)}$ and $C^{(2)}$ are universal. From our results in Fig.~\ref{fig:Im_D1_large_a0_and_a2}, we determine that $C^{(0)} = 12.87(1)$ and $C^{(2)} = 5.89(1)$.

\subsection{Strong $p$-wave interactions}
\label{ssec:results_strong_pwave}

Identical bosons with spin $f = 1$ can form a $p$-wave dimer state with $F_{\mathrm{2b}} = 1$. At resonance, the corresponding $p$-wave scattering volume $v_{1}$ diverges. Near the resonance, the $p$-wave dimer state is quasibound for $v_1<0$ and bound for $v_1>0$. Even though resonant $p$-wave interactions cannot influence the scattering state of three identical spinless bosons, recent work demonstrated that they give rise to a $\sqrt{-v}$ scaling of the three-body scattering hypervolume for dissimilar particles \cite{mestrom2021hypervolumeBBX}, where $v$ is the relevant $p$-wave scattering volume. Similarly, we demonstrate here that $D_1$ diverges as $\sqrt{-v_{1}}$ at resonance. More specifically, we find that
\begin{equation}\label{eq:D1_limit_large_a1}
\begin{aligned}
D_1/\sqrt{-v_1} \underset{|v_{1}| \to \infty}{=} &- \frac{160}{9} \sqrt{6} \pi^2 \left(a_0 - a_2\right)^2
\frac{1}{\sqrt{\tilde{r}_{1}}},
\end{aligned}
\end{equation}
where $\tilde{r}_{1}$ is the $p$-wave effective range corresponding to $V_{F_{\mathrm{2b}} =1}$. This universal result is derived in Appendix~\ref{sec:D1_resonant_p_wave} and is independent of the choice of $\rho_{1}$. It originates from a few dominant three-body scattering processes involving the $p$-wave component of $V_{F_{\mathrm{2b}}=1}$ and the $s$-wave components of $V_{F_{\mathrm{2b}}=0}$ and $V_{F_{\mathrm{2b}}=2}$. For $v_1 \to -\infty$, these scattering processes are elastic and only $\mathrm{Re}\left(D_1\right)$ diverges. For $v_1 \to +\infty$, they are inelastic and describe decay into the shallow $p$-wave dimer state with $F_{\mathrm{2b}} = 1$. In this limit, we expect that $\mathrm{Re}\left(D_1\right)$ diverges logarithmically with respect to $v_1$ in a similar way as was found in Ref.~\cite{mestrom2021hypervolumeBBX}. The divergent behavior of $D_1$ described by Eq.~\eqref{eq:D1_limit_large_a1} could strongly influence spinor condensates as we will see in Section~\ref{sec:spinor_condensates}.



Finally, we note that $D_1$ can only diverge at a $p$-wave dimer resonance when $a_0 \neq a_2$ as can be seen from Eq.~\eqref{eq:D1_limit_large_a1}. This is consistent with the fact that the three-body scattering state for identical spin-1 bosons with $F_{\mathrm{3b}} = 1$ maps onto the one for identical spinless bosons when $V_{F_{\mathrm{2b}} = 0} = V_{F_{\mathrm{2b}} = 2}$ as discussed in Section~\ref{sec:theory_spinor_3body}.

\begin{table}[tbp]
  \centering
  \caption{$S$-wave scattering lengths $a_{F_{\mathrm{2b}}}$ and $p$-wave scattering volumes $v_{F_{\mathrm{2b}}}$ for several bosonic alkali-metal atoms with spin $f= 1$. These values are obtained from coupled-channels calculations using the potentials from Refs.~\cite{julienne2014potentialsLi, knoop2011potentialNa, tiemann2020potentialsK, strauss2010potentialRb}. Our results for $a_0$ and $a_2$ are consistent with the values stated in Refs.~\cite{stamperkurn2013reviewSpinorBoseGas, huh2020spinorBEC7Li, crubellier1999spinorScatLNa23, lysebo2010spinorScatLK41, klausen2001spinorScatLRb87, vankempen2002spinorScatLRb87}. We also give $r_{\mathrm{vdW}}$ in units of the Bohr radius $a_{\mathrm{B}}$.
  }
  \label{tab:data_Jinglun_spinor_atomic species}
    \begin{tabular}{ccccc}
    \toprule
    Element & $a_{0}/r_{\mathrm{vdW}}$ & $a_{2}/r_{\mathrm{vdW}}$ & $v_1/r_{\mathrm{vdW}}^3$ & $r_{\mathrm{vdW}}/a_{\mathrm{B}}$ \\
    \hline
   ${}^{7}$Li & 0.74 & 0.22 & 0.02 & 32.49 \\
   ${}^{23}$Na & 1.09 & 1.21 & 0.46 & 44.96 \\
   ${}^{41}$K & 1.05 & 0.97 & 0.14 & 65.43 \\
   ${}^{87}$Rb & 1.230 & 1.216 & 0.70 & 82.64 \\
    \bottomrule
    \end{tabular}\\
\end{table}

\begin{figure*}[hbtp]
	\begin{subfigure}
	\centering
	\includegraphics[width=3.4in]{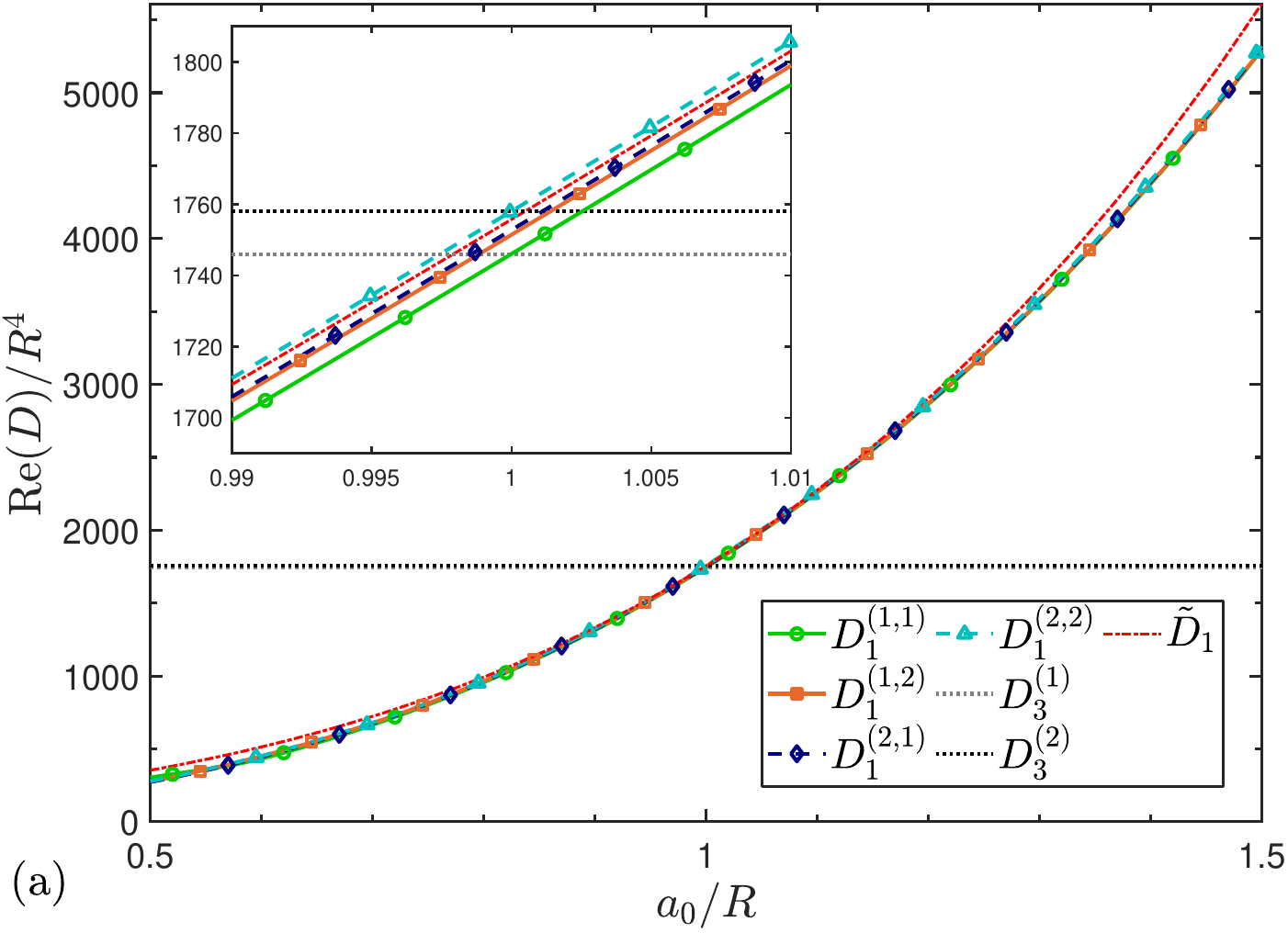}
	\end{subfigure}
\quad
	\begin{subfigure}
	\centering
	\includegraphics[width=3.4in]{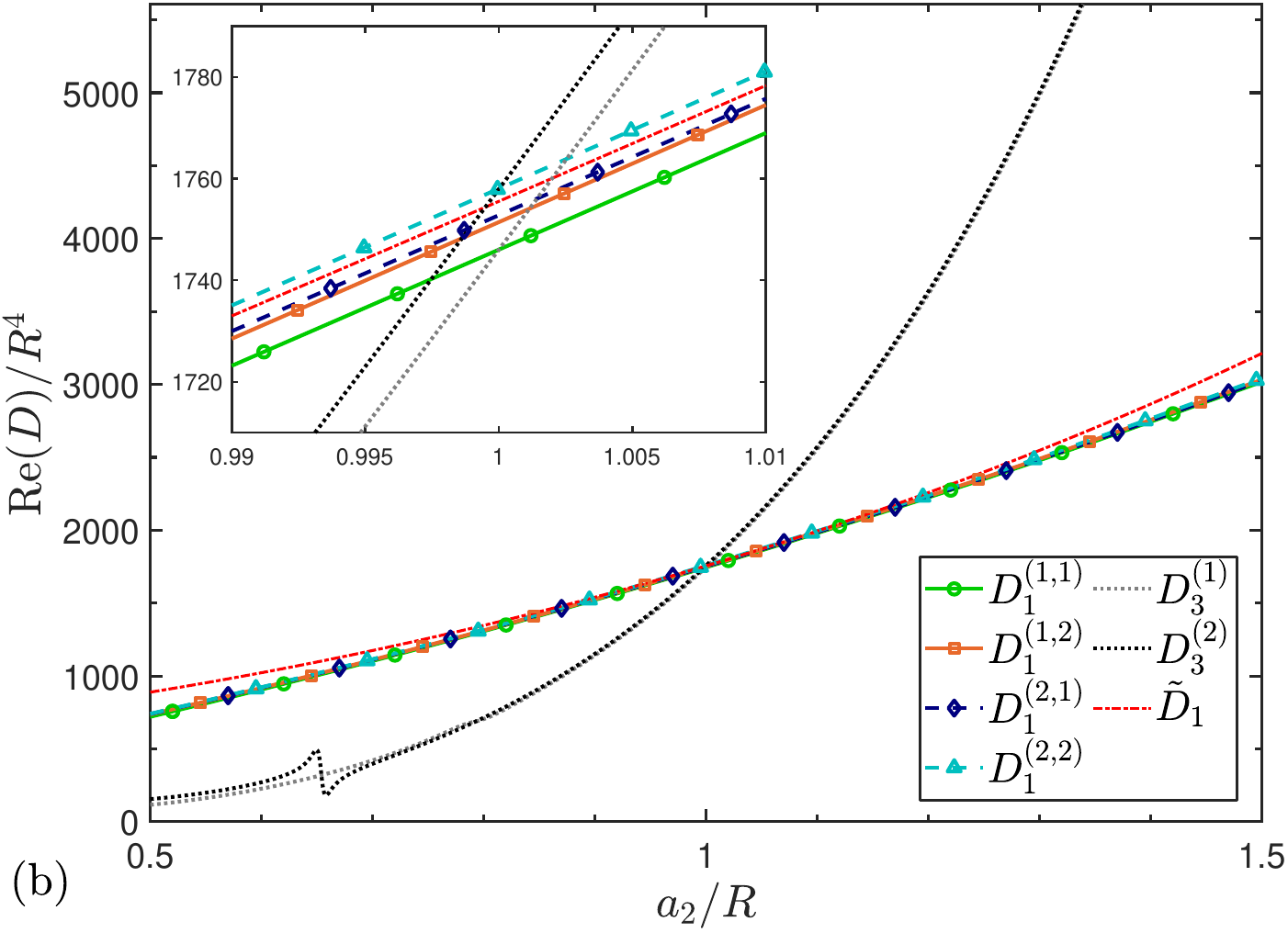}
	\end{subfigure}
	
	\begin{subfigure}
	\centering
	\includegraphics[width=3.4in]{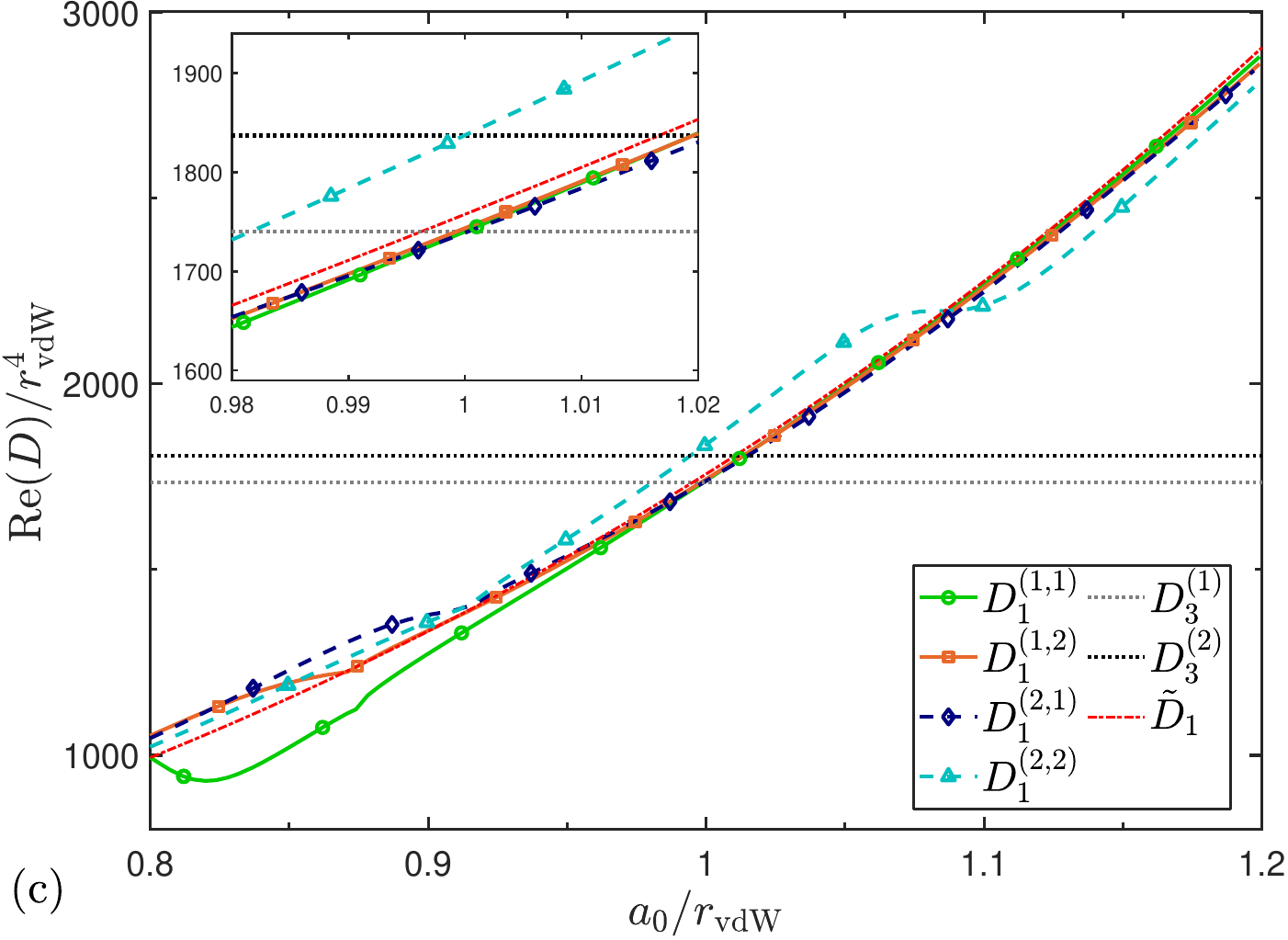}
	\end{subfigure}
\quad
	\begin{subfigure}
	\centering
	\includegraphics[width=3.4in]{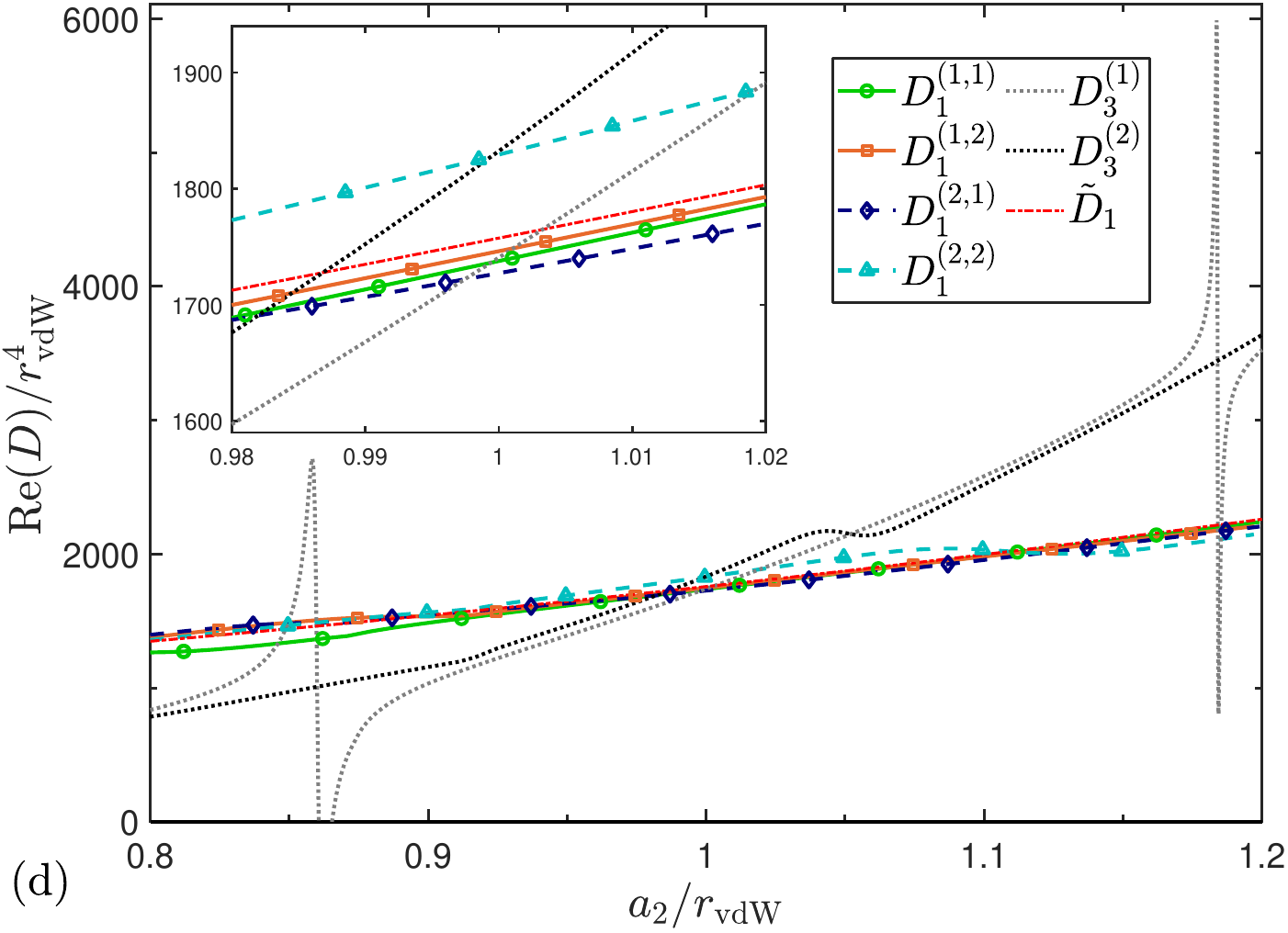}
	\end{subfigure}
    \caption{Real part of the three-body scattering hypervolumes $D_1$ and $D_3$ corresponding to multiple pairwise square-well (a, b) and van der Waals (c, d) potentials that are described by Eq.~\eqref{eq:V_F2b_SqW} and Eq.~\eqref{eq:V_F2b_ZerovdW}, respectively. The range of the potentials $V_{F_{\mathrm{2b}} = 0}$ and $V_{F_{\mathrm{2b}} = 2}$ is identical and is given by $R\equiv R_0 = R_2$ (a, b) or $r_{\mathrm{vdW}}$ (c, d). We fix $a_2/R = 1$ (a), $a_0/R = 1$ (b), $a_2/r_{\mathrm{vdW}} = 1$ (c) or $a_0/r_{\mathrm{vdW}} = 1$ (d) and vary the other scattering length. In all cases, we set $V_{F_{\mathrm{2b}} = 1} = 0$ and $\rho_1 = \rho_3 = |a_2|$. Additionally, we display the curve for $\tilde{D}_1$ defined in Eq.~\eqref{eq:tildeD_vs_tilde_a_and_rho} with $a_{\mathrm{hh},1}^{+}/R = -0.0097$ (a, b) or $a_{\mathrm{hh},1}^{+}/r_{\mathrm{vdW}} = -0.01$ (c, d). In (c, d), there is a $d$-wave dimer resonance associated with $V_{F_{\mathrm{2b}}}$ with $N_{F_{\mathrm{2b}}} = 1$ (2) at $a_{F_{\mathrm{2b}}}/r_{\mathrm{vdW}} = 0.87$ (0.92), resulting in a very small kink in the curves for $\mathrm{Re}\left(D_1\right)$ and $\mathrm{Re}\left(D_3\right)$ \cite{dincao2012dwave}.}
    \label{fig:Dhyp_spinor_SqW_R0=R2_and_ZerovdW_rvdW0=rvdW2}
\end{figure*}

\subsection{Weak interactions}
\label{ssec:results_weak_interactions}

So far, we have studied the behavior of $D_1$ for resonant $s$- and $p$-wave interactions. However, most alkali-metal atoms with $f = 1$ are weakly interacting in the absence of an external magnetic field. This can be seen from Table~\ref{tab:data_Jinglun_spinor_atomic species} which shows the $s$-wave scattering lengths and $p$-wave scattering volumes for several $f = 1$ atoms. For ${}^{23}$Na, ${}^{41}$K and ${}^{87}$Rb, both scattering lengths $a_0$ and $a_2$ are comparable to the van der Waals range $r_{\mathrm{vdW}} = \left(m C_6/\hbar^2\right)^{1/4}/2$ which is the typical length scale corresponding to the van der Waals tail $-C_6/r^6$ of the interatomic interaction. Therefore, we focus our study of $D_1$ on positive $a_0$ and $a_2$ with values around the interaction range.

In our analysis, we consider two different local finite-range potentials to mimic real interatomic interactions. We investigate a simplified square-well interaction model and a more realistic van der Waals interaction potential. Despite their differences, we will see that they both give rise to similar behavior of $\mathrm{Re}\left(D_1\right)$ which is only weakly dependent on finite-range effects in the considered interaction regime. This allows us to apply our results to atomic systems, for which we analyze the effects of three-body spin-mixing collisions on the properties of spinor condensates in Section~\ref{sec:spinor_condensates}. Such an analysis requires knowledge of both $D_1$ and $D_3$. For this reason we also display the curves for $D_3$ in this section.

The considered potentials $V_{F_{\mathrm{2b}}}$ are either square-well potentials with depths $\Omega_{F_{\mathrm{2b}}}$ and ranges $R_{F_{\mathrm{2b}}}$, i.e.,
\begin{equation}\label{eq:V_F2b_SqW}
V_{F_{\mathrm{2b}}}(r)=\begin{cases}-\Omega_{F_{\mathrm{2b}}},&\mbox{$0\leq r<R_{F_{\mathrm{2b}}}$},\\0,&\mbox{$r\geq R_{F_{\mathrm{2b}}}$},\end{cases}
\end{equation}
or the van der Waals potentials
\begin{equation}\label{eq:V_F2b_ZerovdW}
V_{F_{\mathrm{2b}}}(r)= \begin{cases} 0,&\mbox{$0\leq r<\lambda_{F_{\mathrm{2b}}}$},
\\-\frac{C_{6}}{r^6},&\mbox{$r\geq \lambda_{F_{\mathrm{2b}}}$}.\end{cases}
\end{equation}
Here $r$ represents the distance between two particles. For simplicity, we set $V_{F_{\mathrm{2b}} = 1} = 0$. This leaves us with two interaction potentials $V_{F_{\mathrm{2b}} = 0}$ and $V_{F_{\mathrm{2b}} = 2}$ for which we tune $a_0$, $a_2$ and the number of two-body bound states by adjusting $\Omega_0$ and $\Omega_2$ in the case of Eq.~\eqref{eq:V_F2b_SqW} and $\lambda_0$ and $\lambda_2$ in the case of Eq.~\eqref{eq:V_F2b_ZerovdW}. We add additional indices to $D_1$ and $D_3$ to indicate the number of $s$-wave dimer states $N_0$ and $N_2$ that are supported by the potentials $V_{F_{\mathrm{2b}} = 0}$ and $V_{F_{\mathrm{2b}} = 2}$, respectively, i.e., $D_1^{(N_0, N_2)}$ and $D_3^{(N_2)}$.


Figure~\ref{fig:Dhyp_spinor_SqW_R0=R2_and_ZerovdW_rvdW0=rvdW2} shows the behavior of $\mathrm{Re}\left(D_1\right)$ and $\mathrm{Re}\left(D_3\right)$ for positive $a_0$ and $a_2$ when the range of $V_{F_{\mathrm{2b}} = 0}$ and $V_{F_{\mathrm{2b}} = 2}$ is identical (i.e., $R_0 = R_2$ in the case of Eq.~\eqref{eq:V_F2b_SqW}). 
It demonstrates that $D_1$ and $D_3$ are only weakly dependent on the short-range behavior of the potentials in the considered regime. This weak dependence of $\mathrm{Re}\left(D_3\right)$ on the short-range details for $a_2/R_2 \gtrsim 0.7$ and $a_2/r_{\mathrm{vdW}} \gtrsim 0.6$ was already recognized in Ref.~\cite{mestrom2020hypervolumeVdW} which studied the three-body scattering hypervolume for spinless bosons. Their analysis resulted in the universal formula
\begin{equation}\label{eq:D3approx_vs_a2_and_rho3}
\begin{aligned}
\mathrm{Re}\left(D_3\right) \approx& 1689 \left(a_2 - a_{\mathrm{hh},3}^{+}\right)^4 
\\
&+ 64 \pi \left(4 \pi - 3\sqrt{3}\right) a_2^4 \ln|a_2/\rho_3|,
\end{aligned}
\end{equation}
where the parameter $a_{\mathrm{hh},3}^{+}$ accounts effectively for finite-range effects at $a_2>0$. A typical value is $a_{\mathrm{hh},3}^{+}/R_2 = -0.01$ for square-well potentials and $a_{\mathrm{hh},3}^{+}/r_{\mathrm{vdW}} = -0.01$ for van der Waals potentials \cite{mestrom2020hypervolumeVdW}. The imaginary parts of $D_1$ and $D_3$ behave nonuniversally, since three-body recombination into deeply bound dimer states requires the particles to approach each other closely, and are therefore not shown in Fig.~\ref{fig:Dhyp_spinor_SqW_R0=R2_and_ZerovdW_rvdW0=rvdW2}. We note that $\mathrm{Re}\left(D_1\right)$ and $\mathrm{Re}\left(D_3\right)$ are more affected by three-body and $d$-wave two-body resonances for the van der Waals interaction in Eq.~\eqref{eq:V_F2b_ZerovdW} than for the square-well potential in Eq.~\eqref{eq:V_F2b_SqW}.  However, apart from these features the curves for $\mathrm{Re}\left(D_1\right)$ are very similar and we will now describe this universal behavior analytically. 

\begin{figure}[btp]
	\begin{subfigure}
	\centering
	\includegraphics[width=3.4in]{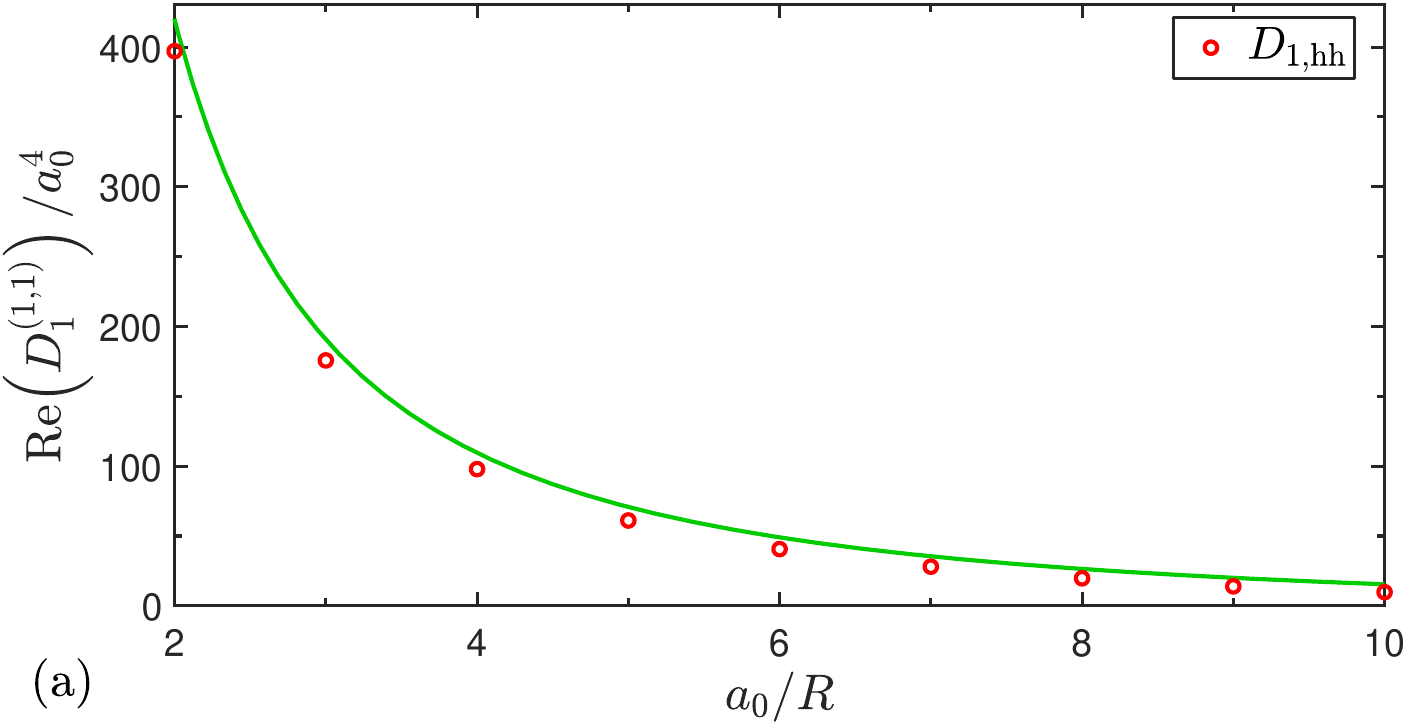}
	\end{subfigure}
\quad
	\begin{subfigure}
	\centering
	\includegraphics[width=3.4in]{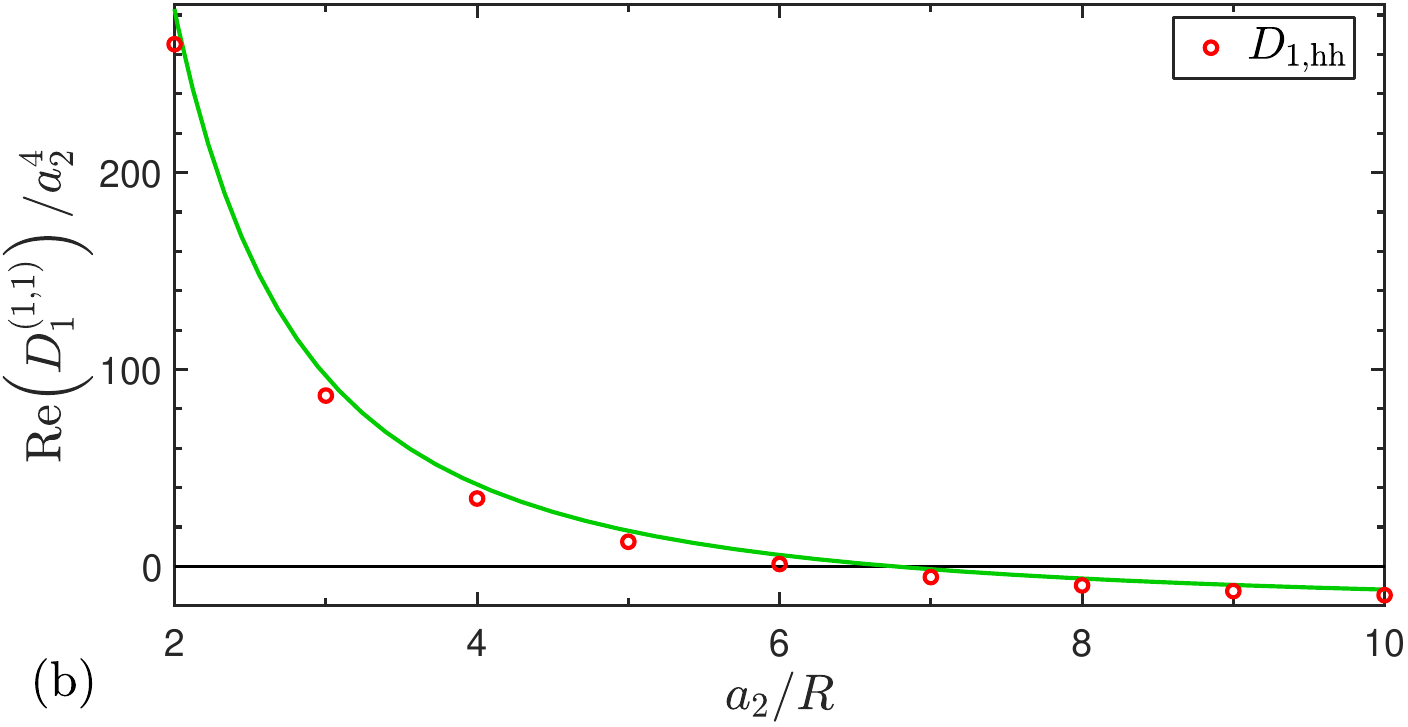}
	\end{subfigure}
    \caption{Real part of the three-body scattering hypervolume $D_1^{(1,1)}$ corresponding to $V_{F_{\mathrm{2b}}}$ in Eq.~\eqref{eq:V_F2b_SqW}. We set $R_0 = R_2 = R$ and $V_{F_{\mathrm{2b}} = 1} = 0$. We consider (a) $2 \leq a_0/R \leq 10$ with $a_2/R = 1$ and $\rho_1 = |a_0|$ and (b) $2 \leq a_2/R \leq 10$ with $a_0/R = 1$ and $\rho_1 = |a_2|$. The circles represent $D_{1,\mathrm{hh}}$ that is determined for various ratios $a_2/a_0$ in the same way as in Fig.~\ref{fig:Dhh_F3b_1_vs_X}(a).}
    \label{fig:D1_SqW_Res_1_1_ScatL_2_10_D1hh}
\end{figure}

The universality of $\mathrm{Re}\left(D_1\right)$ suggests that it can be quantified by a simple formula based on Eq.~\eqref{eq:Dh_hh_F3b_1_Taylor_X} in a similar way as was done for $\mathrm{Re}\left(D_3\right)$ in Eq.~\eqref{eq:D3approx_vs_a2_and_rho3}. In Fig.~\ref{fig:Dhyp_spinor_SqW_R0=R2_and_ZerovdW_rvdW0=rvdW2}, we compare our results for $\mathrm{Re}\left(D_1\right)$ to 
\begin{equation}\label{eq:tildeD_vs_tilde_a_and_rho}
\tilde{D}_1 \equiv 1689 \left(\tilde{a} - a_{\mathrm{hh},1}^{+}\right)^4 + 64 \pi \left(4 \pi - 3\sqrt{3}\right) \tilde{a}^4 \ln|\tilde{a}/\rho_1|,
\end{equation}
which works best near the point $a_0 = a_2$ because of Eq.~\eqref{eq:Dh_hh_F3b_1_Taylor_X}. 
The quantity $\tilde{a}$ was defined in Eq.~\eqref{eq:atilde_def}.
The offset $a_{\mathrm{hh},1}^{+}$ introduces a shift due to finite-range effects of the interaction potentials. Figure~\ref{fig:Dhyp_spinor_SqW_R0=R2_and_ZerovdW_rvdW0=rvdW2} shows that $\tilde{D}_1$ with $a_{\mathrm{hh},1}^{+}/r_{\mathrm{vdW}} \simeq -0.01$ or $a_{\mathrm{hh},1}^{+}/R \simeq -0.01$ and $R \equiv R_0 = R_2$ matches $\mathrm{Re}\left(D_1\right)$ well when $a_0 \approx a_2$. More generally, $\mathrm{Re}\left(D_1\right)$ is well described by the curve for $D_{1,\mathrm{hh}}$ shifted by finite-range effects, even outside the regime where $a_0 \approx a_2$. This is shown in Fig.~\ref{fig:D1_SqW_Res_1_1_ScatL_2_10_D1hh} for both $a_0 > a_2$ and $a_2 > a_0$. This demonstrates that the dominant contribution to $\mathrm{Re}\left(D_1\right)$ comes from hard-hyperspherelike collisions.
We use this finding to make quantitative predictions for atomic systems in Section~\ref{sec:spinor_condensates}.

The small values of $a_{\mathrm{hh},1}^{+}/r_{\mathrm{vdW}}$ and $a_{\mathrm{hh},1}^{+}/R$ indicate that finite-range effects on $\mathrm{Re}\left(D_1\right)$ are small in the interaction regimes considered in Fig.~\ref{fig:Dhyp_spinor_SqW_R0=R2_and_ZerovdW_rvdW0=rvdW2}. These small effects result from $V_{F_{\mathrm{2b}} = 0}$ and $V_{F_{\mathrm{2b}} = 2}$. Additional finite-range effects on $\mathrm{Re}\left(D_1\right)$ arise from $V_{F_{\mathrm{2b}} = 1}$ which we have not incorporated in Fig.~\ref{fig:Dhyp_spinor_SqW_R0=R2_and_ZerovdW_rvdW0=rvdW2}. However, we expect that it is fine to neglect $V_{F_{\mathrm{2b}} = 1}$ as long as it does not have a longer range than $V_{F_{\mathrm{2b}} = 0}$ and $V_{F_{\mathrm{2b}} = 2}$ in which case the effects of $V_{F_{\mathrm{2b}} = 1}$ on $a_{\mathrm{hh},1}^{+}$ should not exceed those of $V_{F_{\mathrm{2b}} = 0}$ and $V_{F_{\mathrm{2b}} = 2}$. 
This argument is based on the fact that the coupling matrix elements $W_{F_{\mathrm{2b}},F_{\mathrm{2b}}'}^{(1)}$ that are defined in Eq.~\eqref{eq:def_W_F2b_F2bprime^(F3b)} and appear in the three-body integral equation \eqref{eq:breve_mathsfA_nl_q_spinor} are of the same order of magnitude as we show in Eq.~\eqref{eq:tildeW_matrix_F3b=1_full}.

As will be discussed in Section~\ref{sec:MF_spinor_BEC}, the difference $\mathrm{Re}\left(D_3 - D_1\right)$ influences the ground state of a spinor BEC. In general, we find that $\mathrm{Re}\left(D_3 - D_1\right) > 0$ for $a_2 - a_0 > 0$ and $\mathrm{Re}\left(D_3 - D_1\right) < 0$ for $a_2 - a_0 < 0$ for the scattering lengths considered in Fig.~\ref{fig:Dhyp_spinor_SqW_R0=R2_and_ZerovdW_rvdW0=rvdW2}. However, close to the point where $a_0 = a_2$ the difference $\mathrm{Re}\left(D_3 - D_1\right)$ is generally small compared to $\mathrm{Re}\left(D_3\right)$ and $\mathrm{Re}\left(D_1\right)$ and the sign and magnitude of this difference depend crucially on the short-range details of $V_{F_{\mathrm{2b}}}$ which are the depths $\Omega_{F_{\mathrm{2b}}}$ for Eq.~\eqref{eq:V_F2b_SqW} and the parameters $\lambda_{F_{\mathrm{2b}}}$ for Eq.~\eqref{eq:V_F2b_ZerovdW}. We note that $D_1^{(1,1)} = D_3^{(1)}$ and $D_1^{(2,2)} = D_3^{(2)}$ at the point $a_0 = a_2$ as can be seen in the insets of Fig.~\ref{fig:Dhyp_spinor_SqW_R0=R2_and_ZerovdW_rvdW0=rvdW2}. This follows from the fact that $V_{F_{\mathrm{2b}} = 0} = V_{F_{\mathrm{2b}} = 2}$ at this point as was discussed in Section~\ref{sec:theory_spinor_3body}.


In Appendix~\ref{sec:SqW_D1_R0_neq_R2} we extend our analysis of $\mathrm{Re}\left(D_1\right)$ for positive $a_0$ and $a_2$ to cases for which $R_0 \neq R_2$ in Eq.~\eqref{eq:V_F2b_SqW}. There we show that finite-range effects become important when $R_0$ or $R_2$ is larger than $a_0$ or $a_2$. The main effect is an overall shift in $\mathrm{Re}\left(D_1\right)$ via $a_{\mathrm{hh},1}^{+}$. This further demonstrates the significance of hard-hyperspherelike collisions in this weakly interacting regime which are well described by Eq.~\eqref{eq:tildeD_vs_tilde_a_and_rho} for $a_0 \approx a_2$.

Our universal description for $\mathrm{Re}\left(D_3\right)$ and $\mathrm{Re}\left(D_1\right)$ in Eqs.~\eqref{eq:D3approx_vs_a2_and_rho3} and \eqref{eq:tildeD_vs_tilde_a_and_rho} can be used to make quantitative predictions for bosonic atoms with $f = 1$. The atoms in Table~\ref{tab:data_Jinglun_spinor_atomic species} have nuclear spin $I = 3/2$ and electron spin $S = 1/2$. Their electronic ground state consists of two hyperfine levels with spin $f = 1$ and 2 which are three- and fivefold degenerate, respectively. The interaction potential between two atoms is not diagonal in the spin quantum numbers $f_1$ and $f_2$ \cite{pethick2002bec}. Therefore, atoms can generally not be treated as spin-1 particles when studying three-body collisions. Exceptions include the universal behavior of the three-body scattering hypervolumes $D_1$ and $D_3$ for resonant $s$- and $p$-wave interactions as presented in Sections~\ref{ssec:results_strong_swave} and \ref{ssec:results_strong_pwave}. In the weakly interacting regime where $a_0 \approx r_{\mathrm{vdW}}$ and $a_2 \approx r_{\mathrm{vdW}}$, we have found that the real parts of $D_1$ and $D_3$ are not much influenced by the short-range behavior of the interaction potential in the absence of trimer resonances. Therefore, we expect that we can still make predictions for $\mathrm{Re}\left(D_1\right)$ and $\mathrm{Re}\left(D_3\right)$ corresponding to ${}^{23}$Na, ${}^{41}$K and ${}^{87}$Rb, for which $a_0 \approx a_2  \approx r_{\mathrm{vdW}}$ (see Table~\ref{tab:data_Jinglun_spinor_atomic species}). For these atoms, we estimate the effects of three-body spin-mixing collisions on the properties of spinor BECs in Section~\ref{sec:spinor_condensates}.

\section{Spinor condensates}
\label{sec:spinor_condensates}

So far we have analyzed the scattering of three identical spin-1 bosons at zero collision energy. Here we study the effect of these three-body collisions on spinor condensates. First, we present a many-body theory of a spin-1 BEC that includes effective three-body interaction strengths defined via $D_1$ and $D_3$. From our results in Section~\ref{sec:results_spin_1}, we identify several regimes where these interaction strengths are important. Furthermore, we estimate the effective three-body interaction strengths for several atomic spinor BECs and discuss some possibilities to increase these interaction strengths. Finally, we analyze how a weak external magnetic field can be used to observe signatures of three-body collisions in the spin-mixing dynamics.




\subsection{Static properties of spin-1 BECs}
\label{sec:MF_spinor_BEC}

Here we consider a many-body theory for a spin-1 BEC that incorporates effective two- and three-body contact interactions. The corresponding interaction strengths are connected to the two-body scattering lengths $a_0$ and $a_2$ and the three-body scattering hypervolumes $D_1$ and $D_3$ which we assume to be real in this section. 
Below we follow the many-body theories presented in Refs.~\cite{ho1998spinorBEC,ohmi1998spinorBEC,law1998spinorBEC, pethick2002bec,kawaguchi2012reviewSpinorBECs,stamperkurn2013reviewSpinorBoseGas} which only considered effective two-body interactions and  Refs.~\cite{mahmud2013spinor3bodyLattice,colussi2014EfimovSpinor,colussi2016EfimovSpinor} which include effective three-body interactions.

We start by introducing the field annihilation operator $\hat{\psi}_{\alpha}(\mathbf{r})$ associated with particles in the spin state $\lvert f = 1, m_f = \alpha \rangle$ where $\alpha = -1$, 0, 1. 
We consider a homogeneous spinor BEC and neglect quantum depletion of the condensate. This leads to $\hat{\psi}_{\alpha}(\mathbf{r}) \approx \hat{a}_{\alpha}/\sqrt{W}$,
where $W$ denotes the spatial volume in which the particles live and $\hat{a}_{\alpha}$ destroys spin-1 bosons in the zero-momentum state with $m_f = \alpha$. The many-body Hamiltonian is then given by 
\begin{widetext}
\begin{eqnarray}\label{eq:Hamiltonian_spin-1_MF_2b_3b_contact_simplified}
\begin{aligned}
\hat{H} =& \frac{2 \pi \hbar^2}{m W} \Bigg(c_{\mathrm{2b}}^{\mathrm{di}} \sum_{\alpha,\beta}  \hat{a}^{\dagger}_{\alpha} \hat{a}^{\dagger}_{\beta} \hat{a}_{\beta} \hat{a}_{\alpha}
+ c_{\mathrm{2b}}^{\mathrm{ex}} \sum_{\alpha,\beta,\alpha',\beta'}\hat{a}^{\dagger}_{\alpha} \hat{a}^{\dagger}_{\beta}
\, \mathbf{f}_{\alpha \alpha'}\cdot \mathbf{f}_{\beta \beta'} \,
\hat{a}_{\beta'} \hat{a}_{\alpha'}
\Bigg)
+ \frac{\hbar^2}{6 m W^2} \Bigg(c_{\mathrm{3b}}^{\mathrm{di}} \sum_{\alpha,\beta,\gamma}  \hat{a}^{\dagger}_{\alpha} \hat{a}^{\dagger}_{\beta} \hat{a}^{\dagger}_{\gamma} \hat{a}_{\gamma} \hat{a}_{\beta} \hat{a}_{\alpha}
\\
&+ c_{\mathrm{3b}}^{\mathrm{ex}} \sum_{\alpha,\beta,\gamma,\alpha',\beta',\gamma'}\hat{a}^{\dagger}_{\alpha} \hat{a}^{\dagger}_{\beta} 
 \hat{a}^{\dagger}_{\gamma} 
 \left(
\mathbf{f}_{\alpha \alpha'}\cdot \mathbf{f}_{\beta \beta'} 
+
\mathbf{f}_{\beta \beta'}\cdot \mathbf{f}_{\gamma \gamma'}
+
\mathbf{f}_{\gamma \gamma'}\cdot \mathbf{f}_{\alpha \alpha'}
\right)
\hat{a}_{\gamma'} \hat{a}_{\beta'} \hat{a}_{\alpha'}
\Bigg).
\end{aligned}
\end{eqnarray}
\end{widetext}
Here $\mathbf{f}_{\alpha \alpha'}$ are the matrix components of the single-particle spin vector $\mathbf{f}$ in the $\lvert f = 1, m_f \rangle$ basis.
The coefficients $c_{\mathrm{2b}}^{\mathrm{di}}$, $c_{\mathrm{2b}}^{\mathrm{ex}}$, $c_{\mathrm{3b}}^{\mathrm{di}}$ and $c_{\mathrm{3b}}^{\mathrm{ex}}$ are the effective interaction strengths. The ``di" superscript indicates the direct part of the interaction. The ``ex" superscript indicates
spin-exchange processes which include spin-mixing collisions described by $\hat{a}^{\dagger}_{1} \hat{a}^{\dagger}_{-1} \hat{a}_{0} \hat{a}_{0}$, $\hat{a}^{\dagger}_{0} \hat{a}^{\dagger}_{0} \hat{a}_{1} \hat{a}_{-1}$, $\hat{a}^{\dagger}_{\alpha} \hat{a}^{\dagger}_{1} \hat{a}^{\dagger}_{-1} \hat{a}_{\alpha} \hat{a}_{0} \hat{a}_{0}$ and $\hat{a}^{\dagger}_{\alpha} \hat{a}^{\dagger}_{0} \hat{a}^{\dagger}_{0} \hat{a}_{\alpha} \hat{a}_{1} \hat{a}_{-1}$ where $\alpha =-1$, 0 or 1. By defining the number operator $\hat{N} = \sum_{\alpha} \hat{a}^{\dagger}_{\alpha}  \hat{a}_{\alpha}$ and the spin operator $\hat{\mathbf{F}} = \sum_{\alpha,\alpha'} \mathbf{f}_{\alpha \alpha'} \hat{a}^{\dagger}_{\alpha} \hat{a}_{\alpha'}$, the Hamiltonian in Eq.~\eqref{eq:Hamiltonian_spin-1_MF_2b_3b_contact_simplified} can be rewritten as
\begin{equation}\label{eq:H_spin-1_MF_2b_3b_contact_homogeneous_v1}
\begin{aligned}
&\hat{H} = \frac{2 \pi \hbar^2}{m W} \Bigg[c_{\mathrm{2b}}^{\mathrm{di}} \hat{N} \left(\hat{N} - 1\right)
+ c_{\mathrm{2b}}^{\mathrm{ex}} \left(\hat{\mathbf{F}}^2 - 2 \, \hat{N}\right)
\Bigg]
\\
&+ \frac{\hbar^2}{6 m W^2} (\hat{N} -2) \Bigg[c_{\mathrm{3b}}^{\mathrm{di}}  \hat{N} (\hat{N} - 1)
+ 3 c_{\mathrm{3b}}^{\mathrm{ex}} (\hat{\mathbf{F}}^2  - 2 \hat{N})
\Bigg].
\end{aligned}
\end{equation}
Equation~\eqref{eq:H_spin-1_MF_2b_3b_contact_homogeneous_v1} generalizes the many-body theory for spin-1 condensates presented in Ref.~\cite{law1998spinorBEC} by including effective three-body interactions and was previously derived in Ref.~\cite{mahmud2013spinor3bodyLattice}. Following Refs.~\cite{law1998spinorBEC,mahmud2013spinor3bodyLattice}, the corresponding ground-state energy $E_0$ is thus given by 
\begin{equation}\label{eq:energy_spinor_BEC}
\begin{aligned}
E_0 = &\frac{2 \pi \hbar^2}{m W} \Bigg[c_{\mathrm{2b}}^{\mathrm{di}}  N (N -1) + c_{\mathrm{2b}}^{\mathrm{ex}}  \left(F(F+1) - 2 N\right) \Bigg]
\\
&+ \frac{\hbar^2}{6 m W^2} (N-2) \Bigg[c_{\mathrm{3b}}^{\mathrm{di}} N (N-1) 
\\
&+ 3 c_{\mathrm{3b}}^{\mathrm{ex}} \left(F(F+1) - 2 N\right)\Bigg],
\end{aligned}
\end{equation}
where $F$ and $N$ are the total spin and particle number, respectively. We take $N \gg 1$ and define the number density $n = N/W$. When $4 \pi c_{\mathrm{2b}}^{\mathrm{ex}} + c_{\mathrm{3b}}^{\mathrm{ex}} n < 0$, the interaction is ferromagnetic and the ground state has the maximal spin $F = N$. When $4 \pi c_{\mathrm{2b}}^{\mathrm{ex}} + c_{\mathrm{3b}}^{\mathrm{ex}} n > 0$, the interaction is antiferromagnetic and the ground state has the minimal spin $F = 0$ if $N$ is even and $F = 1$ if $N$ is odd \cite{law1998spinorBEC,kawaguchi2012reviewSpinorBECs}.


The effective two-body interaction strengths $c_{\mathrm{2b}}^{\mathrm{di}}$ and $c_{\mathrm{2b}}^{\mathrm{ex}}$ are determined by the scattering lengths $a_0$ and $a_2$ via $c_{\mathrm{2b}}^{\mathrm{di}} = (a_0 + 2 a_2)/3$ and $c_{\mathrm{2b}}^{\mathrm{ex}} = (a_2 - a_0)/3$ \cite{ho1998spinorBEC,ohmi1998spinorBEC, pethick2002bec,kawaguchi2012reviewSpinorBECs,stamperkurn2013reviewSpinorBoseGas}. Similarly, the effective three-body interaction strengths $c_{\mathrm{3b}}^{\mathrm{di}}$ and $c_{\mathrm{3b}}^{\mathrm{ex}}$ can be connected to $D_1$ and $D_3$ via $c_{\mathrm{3b}}^{\mathrm{di}} = (3 D_1 + 2 D_3)/5$ and $c_{\mathrm{3b}}^{\mathrm{ex}} = (D_3 - D_1)/5$ \cite{colussi2014EfimovSpinor,colussi2016EfimovSpinor}. This connection is not unique for two reasons. First of all, the definition of $D_1$ and $D_3$ still depends on the choice for $\rho_1$ and $\rho_3$. Secondly, we could have chosen to replace $D_1$ and $D_3$ in our definitions of $c_{\mathrm{3b}}^{\mathrm{di}}$ and $c_{\mathrm{3b}}^{\mathrm{ex}}$ by $m \hbar^4 \mathcal{U}_0^{(1)}$ and $m \hbar^4 \mathcal{U}_0^{(3)}$, respectively, where $\mathcal{U}_0^{(F_{\mathrm{3b}})}$ is connected to $D_{F_{\mathrm{3b}}}$ via Eq.~\eqref{eq:def_D_F3b_spinor}. However, our choice for $D_1$ and $D_3$ is based on Ref.~\cite{tan2008hypervolume}, which found the three-body scattering hypervolume to be a suitable parameter for the three-body effective interaction of a spinless BEC. These two complications vanish in the limit $|a_0|,|a_2| \to 0$ where $c_{\mathrm{3b}}^{\mathrm{di}}$ and $c_{\mathrm{3b}}^{\mathrm{ex}}$ are uniquely defined. 

Spin exchange via three-body collisions dominates over two-body spin-exchange collisions when $n>n_{\mathrm{c}}$, where we define the critical density $n_{\mathrm{c}} \equiv 4 \pi |c_{\mathrm{2b}}^{\mathrm{ex}}|/|c_{\mathrm{3b}}^{\mathrm{ex}}|$.
This condition is trivially fulfilled for $a_0 = a_2$ in which case $c_{\mathrm{2b}}^{\mathrm{ex}}$ vanishes, whereas $c_{\mathrm{3b}}^{\mathrm{ex}}$ is generally nonzero due to finite-range effects of $V_{F_{\mathrm{2b}}}$.
Other possibilities involve resonant $s$- and $p$-wave two-body interactions for which $c_{\mathrm{3b}}^{\mathrm{ex}}$ diverges as discussed in Section~\ref{sec:results_spin_1}. In particular, strong $p$-wave interactions can make the condensate antiferromagnetic, since Eq.~\eqref{eq:D1_limit_large_a1} implies that $c_{\mathrm{3b}}^{\mathrm{ex}} \to +\infty$ for $v_1 \to -\infty$.
Near trimer resonances $|c_{\mathrm{3b}}^{\mathrm{ex}}|$ can also be significantly enhanced.

\begin{figure}[btp]
	\centering
	\includegraphics[width=3.4in]{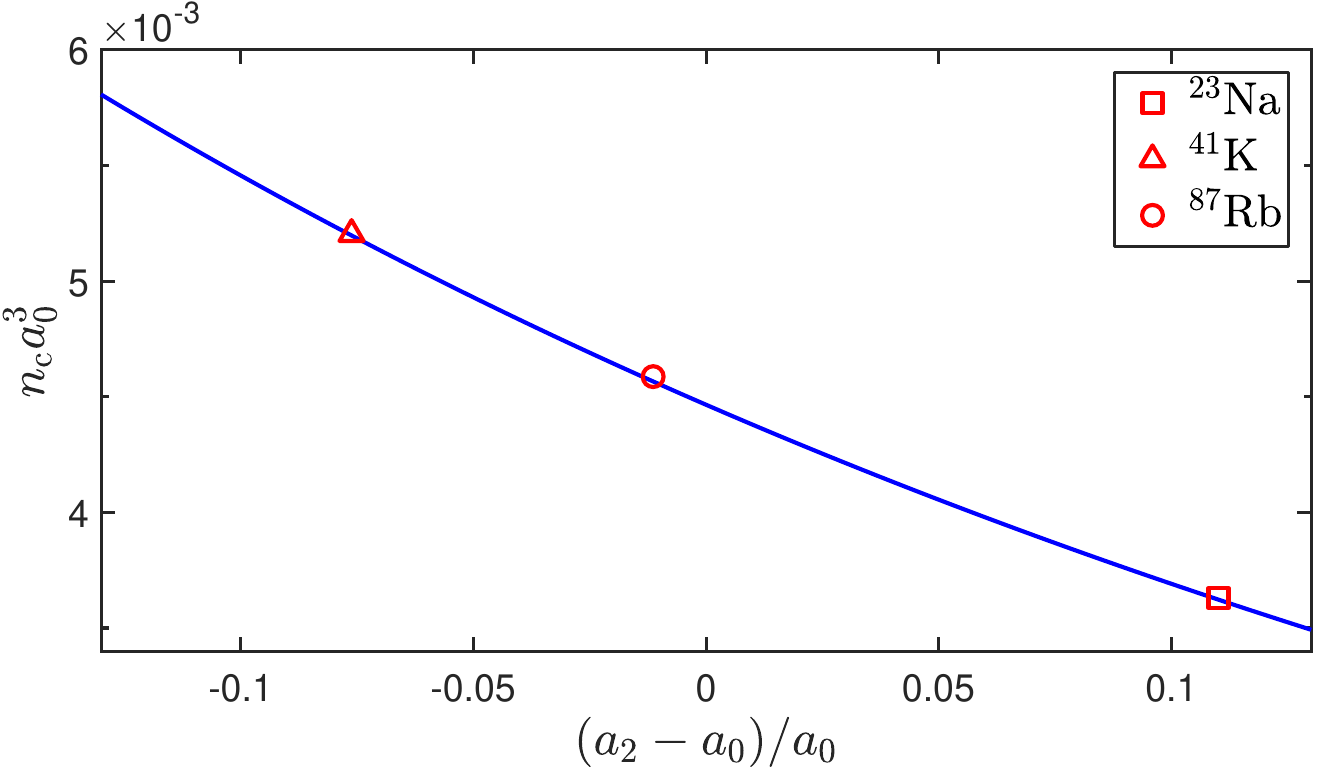}
    \caption{Estimates of the critical density $n_{\mathrm{c}}$ as a function of $(a_2 - a_0)/a_0$. We take $\rho_1 = \rho_3 = |a_2|$ and estimate $D_3 = 1689 (a_2 - a_{\mathrm{hh},3}^{+})^4$ and $D_1 = \tilde{D}_1$. We take $a_0/r_{\mathrm{vdW}} = 1$ and $a_{\mathrm{hh},1}^{+} = a_{\mathrm{hh},3}^{+} = -0.01~r_{\mathrm{vdW}}$ for the blue line. The markers indicate the estimated values for specific spin-1 atoms when taking $a_{\mathrm{hh},1}^{+} = a_{\mathrm{hh},3}^{+} = -0.01~r_{\mathrm{vdW}}$.}
    \label{fig:nCrit_ZeroRange_estimate}
\end{figure}

Taking $\rho_1 = \rho_3 = |a_2|$, we use Eqs.~\eqref{eq:D3approx_vs_a2_and_rho3} and \eqref{eq:tildeD_vs_tilde_a_and_rho} with $a_{\mathrm{hh},1}^{+} = a_{\mathrm{hh},3}^{+}$ to estimate $D_3$ and $D_1 \approx \tilde{D}_1$.
Figure~\ref{fig:nCrit_ZeroRange_estimate} shows how the critical density $n_{\mathrm{c}} = 20 \pi |a_2 -a_0|/3|D_3 - D_1|$ defined above behaves near $a_0 \approx a_2$ under these estimations. At $a_0 = a_2$, our estimate fails because $n_{\mathrm{c}}$ should be zero due to finite-range effects (i.e., $a_{\mathrm{hh},1}^{+} \neq a_{\mathrm{hh},3}^{+}$). Therefore, we cannot make reliable predictions for $n_{\mathrm{c}}$ of ${}^{87}$Rb. For ${}^{23}$Na and ${}^{41}$K, our estimates in Fig.~\ref{fig:nCrit_ZeroRange_estimate} give $n_{\mathrm{c}} \approx 10^{17}~\mathrm{cm}^{-3}$. This density is larger than typical experimental values on the order of $10^{14}~\mathrm{cm}^{-3}$. This demonstrates that three-body spin exchange is usually not important for these experiments and can safely be neglected. Furthermore, our estimates imply that $\mathrm{Re}\left(D_3 - D_1\right) > 0$ for $a_2 - a_0 > 0$ and $\mathrm{Re}\left(D_3 - D_1\right) < 0$ for $a_2 - a_0 < 0$, so that the (anti)ferromagnetic character of a spinor BEC is not changed by three-body collisions. However, this conclusion does not hold for $a_0 = a_2$. 

Calculating the sign and magnitude of $\mathrm{Re}\left(D_3 - D_1\right)$ for atoms is challenging, in particular at $a_0 = a_2$  due to its sensitivity to finite-range effects. Furthermore, the $f = 2$ hyperfine level needs to be included in such calculations, which increases the three-particle Hilbert space significantly. 
We note that new methods which could tackle this problem have been recently developed \cite{secker2021mappedgrid3body,secker2021efimovK39}.

In order to observe strong three-body spin-exchange effects on atomic spinor condensates, the interatomic interactions can be tuned to $a_0 = a_2$ or to an $s$- or $p$-wave dimer resonance as mentioned above. This might be possible via optical \cite{hamley2009opticalFRspin1BEC}, microwave-induced \cite{papoular2010microwaveFR} or radio-frequency-induced \cite{ding2017radiofrequencyFR} Feshbach resonances. In particular, Ref.~\cite{hamley2009opticalFRspin1BEC} identified several photoassociation laser light frequencies for which $a_0 = a_2$ in a ${}^{87}$Rb spin-1 BEC. However, the corresponding losses make experiments in this regime challenging.
Magnetic Feshbach resonances \cite{chin2010feshbach} could be used as well if such resonances are found at extremely low magnetic fields, so that the single-particle spin $f$ can still be treated as a good quantum number.


Finally, we note that atoms also interact via the magnetic dipole-dipole interaction (MDDI) which competes with the spin-exchange interaction, induces new quantum phases in spinor BECs and affects their spin-mixing dynamics \cite{yi2004spinorMDDI,yi2006spinorMDDItrap,yi2006spinorMDDIweakBfield,stamperkurn2013reviewSpinorBoseGas}. However, the effective interaction strength of the MDDI can be tuned for trapped spinor condensates by changing the trap geometry \cite{yi2004spinorMDDI,stamperkurn2013reviewSpinorBoseGas}, so that the MDDI can be effectively turned off.

\subsection{Spin-mixing dynamics}
\label{ssec:spin_dynamics}

When $n \ll n_{\mathrm{c}}$, the spin-mixing dynamics are predominantly determined by two-body spin-mixing collisions. However, by applying an external magnetic field it is possible to cancel this dominant contribution in certain observables describing the spin-mixing dynamics and to see effects of three-body collisions. To illustrate this, we consider a spinor condensate with initially all particles in the $m_f = 0$ state and analyze the dynamics of the spin state populations $N_{m_f}$.
For this purpose, we consider only the part of $\hat{H}$ that contributes to the spin-mixing dynamics. We call this part $\hat{H}_{\mathrm{dyn}}$. It does not involve the linear Zeeman shift, since it is proportional to the magnetization that stays equal to zero, and the direct part of the interaction that is quantified by $c_{\mathrm{2b}}^{\mathrm{di}}$ and $c_{\mathrm{3b}}^{\mathrm{di}}$ \cite{evrard2021spinorBECdynamicsNa23}. Instead, $\hat{H}_{\mathrm{dyn}}$ contains the quadratic Zeeman shift and the spin-exchange interactions. Since the particle number $N_0$ is much larger than $N_{\pm 1}$, we can simplify the Hamiltonian $\hat{H}_{\mathrm{dyn}}$ by retaining only the terms that are quadratic in $\hat{a}_{\pm 1}$ and $\hat{a}_{\pm 1}^{\dagger}$ and by replacing $\hat{a}_{0}$ and $\hat{a}_{0}^{\dagger}$ by $\sqrt{N_0}$. This Bogoliubov approximation leads to \cite{mias2008spinorBECdynamicsFerro,kawaguchi2012reviewSpinorBECs,evrard2021spinorBECdynamicsNa23}
\begin{equation}\label{eq:Hamiltonian_spin_dynamics}
\begin{aligned}
\hat{H}_{\mathrm{dyn}} \approx& \left(q_{\mathrm{Z}} + U_s\right) \left(\hat{a}_{1}^{\dagger} \hat{a}_{1} + \hat{a}_{-1}^{\dagger} \hat{a}_{-1} \right)
\\
&+ U_s \left(\hat{a}_{1}^{\dagger} \hat{a}_{-1}^{\dagger} + \hat{a}_{1} \hat{a}_{-1} \right),
\end{aligned}
\end{equation}
where $q_{\mathrm{Z}}$ is the quadratic Zeeman energy \cite{kawaguchi2012reviewSpinorBECs, stamperkurn2013reviewSpinorBoseGas} and $U_s =  \left(4 \pi n_0 c_{\mathrm{2b}}^{\mathrm{ex}} + n_0^2 c_{\mathrm{3b}}^{\mathrm{ex}}\right) \hbar^2/m$ with number density $n_0 = N_0/W$.
This definition of $U_s$ extends the one defined in Ref.~\cite{evrard2021spinorBECdynamicsNa23} to include the three-body term. The Hamiltonian in Eq.~\eqref{eq:Hamiltonian_spin_dynamics} can be diagonalized using the Bogoliubov transformation \cite{bogoliubov1947superfluidity,pethick2002bec}. The corresponding Bogoliubov excitation energy $\varepsilon_{\mathrm{B}}$ is given by \cite{mias2008spinorBECdynamicsFerro,cui2008spinorBECdynamics,kawaguchi2012reviewSpinorBECs,evrard2021spinorBECdynamicsNa23}
\begin{equation}
\varepsilon_{\mathrm{B}} = \sqrt{q_{\mathrm{Z}} (q_{\mathrm{Z}} + 2 U_s)}.
\end{equation}

Considering positive $q_{\mathrm{Z}}$, the initial phase of the spinor condensate is stable for $U_s > -q_{\mathrm{Z}}/2$. When $c_{\mathrm{2b}}^{\mathrm{ex}}<0$, the magnetic field can be tuned to the value at which 
\begin{equation}\label{eq:qZ_vs_c3b1}
q_{\mathrm{Z}} = -\frac{8 \pi \hbar^2 n_0 c_{\mathrm{2b}}^{\mathrm{ex}}}{m},
\end{equation}
so that 
\begin{equation}\label{eq:Bogoliubov_energy_critical}
\begin{aligned}
\varepsilon_{\mathrm{B}} &= \hbar n_0 \sqrt{2 q_{\mathrm{Z}} c_{\mathrm{3b}}^{\mathrm{ex}}/m} 
\\
&= \frac{4 \hbar^2 n_0^{3/2} \sqrt{-\pi c_{\mathrm{2b}}^{\mathrm{ex}} c_{\mathrm{3b}}^{\mathrm{ex}}}}{m}.
\end{aligned}
\end{equation} 
The $n_0^{3/2}$ scaling of $\varepsilon_{\mathrm{B}}$ provides a clear signature of three-body spin mixing. The Bogoliubov frequency $\varepsilon_{\mathrm{B}}/\hbar$ can be measured by tracking the average number of pairs with $m_f = \pm 1$, i.e., $N_p = (N_{1} + N_{-1})/2$, as a function of time. When $N_p = 0$ at time $t = 0$, $N_p$ oscillates as $N_p(t) = (U_s/\varepsilon_{\mathrm{B}})^2 \sin^2(\varepsilon_{\mathrm{B}} t/\hbar)$ for $c_{\mathrm{3b}}^{\mathrm{ex}} > 0$ \cite{cui2008spinorBECdynamics,evrard2021spinorBECdynamicsNa23}, whereas it grows exponentially as $N_p(t) = |U_s/\varepsilon_{\mathrm{B}}|^2 \sinh^2(|\varepsilon_{\mathrm{B}}| t/\hbar)$ for $c_{\mathrm{3b}}^{\mathrm{ex}} < 0$ at small $t$ for which $N_0 \gg N_p$ \cite{mias2008spinorBECdynamicsFerro, saito2007spinorBECquench}. 
For $c_{\mathrm{3b}}^{\mathrm{ex}} > 0$, our initial assumption $N_0 \gg N_p$ requires that
\begin{equation}\label{eq:N0_gg_Np}
N_0 \gg \frac{m U_s^2}{2 \hbar^2 q_{\mathrm{Z}} n_0^2 c_{\mathrm{3b}}^{\mathrm{ex}}}. 
\end{equation}
When $c_{\mathrm{2b}}^{\mathrm{ex}}>0$, the quadratic Zeeman shift cannot cancel out the two-body spin-exchange term in $\varepsilon_{\mathrm{B}}$. The resulting Bogoliubov oscillations in $N_p(t)$ have recently been observed using a spin-1 condensate of ${}^{23}$Na atoms \cite{evrard2021spinorBECdynamicsNa23}.


To make quantitative predictions of $\varepsilon_{\mathrm{B}}$ in Eq.~\eqref{eq:Bogoliubov_energy_critical} we again use our results of Section~\ref{ssec:results_weak_interactions}. First, we consider a spinor condensate of ${}^{87}$Rb atoms occupying the $m_f = 0$ state. Taking a typical particle density $n_0 = 10^{14}~\mathrm{cm}^{-3}$, Eq.~\eqref{eq:qZ_vs_c3b1} demands that the magnetic field is tuned to $0.29~$G where $q_{\mathrm{Z}}/k_{\mathrm{B}} = 0.29~$nK.
If $|\mathrm{Re}\left(D_3 - D_1\right)| = 100~r_{\mathrm{vdW}}^4$, then Eq.~\eqref{eq:N0_gg_Np} requires $N_0 \gg 87$ and $|\varepsilon_{\mathrm{B}}|/h = 0.32$~Hz. Fortunately, the corresponding timescale is smaller than the typical lifetime $\tau_{\mathrm{life}} = 1/L_3 n_0^2 = 10$~s where we estimated the three-body loss rate $L_3$ to be $10^{-29}~\mathrm{cm}^{6}/$s \cite{burt1997recombinationBECvsThermal, tolra2004recombinationRb87, wolf2019statetostateChemistry, deng2021dipolar, noteLossRateRb87}. For a ${}^{41}$K condensate with $n_0 = 10^{14}~\mathrm{cm}^{-3}$, Eq.~\eqref{eq:qZ_vs_c3b1} is satisfied for $q_{\mathrm{Z}}/k_{\mathrm{B}} = 2.7~$nK which happens at a magnetic field of $0.17~$G. Our simple estimates that we used in Fig.~\ref{fig:nCrit_ZeroRange_estimate} suggest that $\mathrm{Re}\left(D_3 - D_1\right) \approx -373~r_{\mathrm{vdW}}^4$, so that $|\varepsilon_{\mathrm{B}}|/h \approx 1.7$~Hz. We note that higher-order corrections, including the Lee-Huang-Yang \cite{lee1957LHYprogressreport, lee1957LHY} and Wu \cite{wu1959energyBEC, sawada1959energyBEC, hugenholtz1959energyBEC} terms, need to be considered as well in the spin-exchange term for ${}^{41}$K for which $|a_2 - a_0|/a_0$ is only 0.076. However, effects of these corrections on $|\varepsilon_{\mathrm{B}}|$ can be canceled out by properly tuning $q_{\mathrm{Z}}$ in a similar way as we propose to cancel out the effect of $c_{\mathrm{2b}}^{\mathrm{ex}}$.

Finally, we note that the exponential growth of $N_p(t)$ below the critical value of $q_{\mathrm{Z}}$ in Eq.~\eqref{eq:qZ_vs_c3b1} has recently been observed for a spin-1 BEC of ${}^{7}$Li atoms \cite{huh2020spinorBEC7Li}. Tuning the magnetic field strength closer to the critical value could also reveal signatures of three-body spin exchange for these atoms. 

\section{Conclusion}
\label{sec:conclusion_spinor}

We have studied zero-energy scattering of three identical bosons with spin $f =1$ interacting via short-range pairwise potentials. The corresponding three-body scattering hypervolumes $D_1$ and $D_3$ determine the effective three-body interaction strengths in a many-body theory of a spin-1 BEC as well as the three-body recombination rates. $D_3$ maps onto the three-body scattering hypervolume for spinless bosons which has been investigated in previous studies for finite-range potentials \cite{tan2008hypervolume,tan2017hypervolume,mestrom2019hypervolumeSqW,mestrom2020hypervolumeVdW}. We have quantified $D_1$ for resonant $s$- and $p$-wave interactions and for weak interactions. At large $s$-wave scattering lengths $a_0$ and $a_2$, we have derived a first-order Taylor approximation around $a_0 = a_2$ for the part of $D_1$ that characterizes hard-hyperspherelike collisions. In the absence of trimer resonances, this contribution also dominates $\mathrm{Re}\left(D_1\right)$ for positive $a_0$ and $a_2$ with values around the interaction range. We have used this universal description to make quantitative predictions for the effective three-body interaction strengths of several alkali-metal atoms. For ${}^{23}$Na and ${}^{41}$K, we predict that spin mixing via three-body collisions dominates over two-body spin-mixing processes for densities $n\gtrsim 10^{17}~\mathrm{cm}^{-3}$, but it does not change the magnetic nature of the condensate. For ${}^{87}$Rb, this critical density is expected to be lower, but calculations with realistic interaction models are needed for more accurate predictions. By applying a small magnetic field on a ${}^{41}$K or ${}^{87}$Rb condensate, we predict that it is possible to observe clear signatures of three-body spin-mixing processes in the dynamics of the spin state populations.

We have also identified several regimes in which the magnetic properties of a spin-1 BEC strongly depend on the three-body spin-exchange term. Firstly, the spin-mixing dynamics near $a_0 = a_2$ can be dominated by three-body collisions due to finite-range effects. Secondly, strong $s$-wave interactions give rise to large values for both the real and imaginary part of $D_1$ and $D_3$. Thirdly, $D_1$ diverges as $\sqrt{-v_1}$ near a $p$-wave dimer resonance with $F_{\mathrm{2b}} = 1$. Resonant $p$-wave interactions can thus be used to generate a strong antiferromagnetic three-body spin-exchange interaction with a relatively low three-body recombination rate. 
For atomic spinor condensates, these three regimes could be probed by tuning the effective interaction strengths with electromagnetic radiation.

The many-body description of spin-1 BECs considered in this paper could also be extended by including the MDDI. The competition between this long-range interaction and the short-range spin-exchange interactions might result in new phases whose properties are sensitive to three-body interactions. Furthermore, the number of fully symmetric three-body spin channels increases for particles with spin $f>1$. The corresponding three-body scattering processes could lead to novel effects on spin-$f$ condensates.

\section*{Acknowledgements}

We thank Denise Ahmed-Braun, Gijs Groeneveld, and Silvia Musolino for stimulating discussions. This research is financially supported by the Netherlands Organisation for Scientific Research (NWO) under Grant No. 680-47-623. V.E.C. acknowledges additional financial support from Provincia Autonoma di Trento and the Italian MIUR under the PRIN2017 projectCEnTraL.

\newpage
\appendix

\section{Three-body transition amplitude of the spinor system}
\label{sec:U00_amplitude_spinor}

In this appendix we complete the description of the three-body transition amplitude in Eq.~\eqref{eq:U00_momentum_Phi3b} by specifying the coefficients $A_{F_{\mathrm{3b}}}$, $B_{F_{\mathrm{3b}}}$ and $C_{F_{\mathrm{3b}}}$. Following the procedure of Refs.~\cite{mestrom2019hypervolumeSqW,mestrom2021hypervolumeBBX}, we find
\begin{equation}
\begin{aligned}
A_{F_{\mathrm{3b}}} =& -\frac{1}{4 \pi^4} \frac{1}{m \hbar^2} \sum_{F_{\mathrm{2b}}} \left(Z_{F_{\mathrm{2b}},\Sigma_{\mathrm{3b}}}\right)^* a_{F_{\mathrm{2b}}}
\\
& \sum_{F_{\mathrm{2b}}'} a_{F_{\mathrm{2b}}'} W_{F_{\mathrm{2b}}, F_{\mathrm{2b}}'}^{(F_{\mathrm{3b}})}  Z_{F_{\mathrm{2b}}',\Sigma_{\mathrm{3b}}},
\end{aligned}
\end{equation}
\begin{equation}
\begin{aligned}
B_{F_{\mathrm{3b}}} =& \frac{1}{12 \pi^3} \frac{1}{m \hbar^3} \sum_{F_{\mathrm{2b}}} \left(Z_{F_{\mathrm{2b}},\Sigma_{\mathrm{3b}}}\right)^*
\Bigg(a_{F_{\mathrm{2b}}} 
\\
&\sum_{F_{\mathrm{2b}}'} W_{F_{\mathrm{2b}},F_{\mathrm{2b}}'}^{(F_{\mathrm{3b}})} a_{F_{\mathrm{2b}}'} 
\sum_{F_{\mathrm{2b}}''} a_{F_{\mathrm{2b}}''} W_{F_{\mathrm{2b}}', F_{\mathrm{2b}}''}^{(F_{\mathrm{3b}})}  Z_{F_{\mathrm{2b}}'',\Sigma_{\mathrm{3b}}}
\\
& -\frac{3 \sqrt{3}}{2 \pi} a_{F_{\mathrm{2b}}}^2 \sum_{F_{\mathrm{2b}}'} a_{F_{\mathrm{2b}}'}
W_{F_{\mathrm{2b}}, F_{\mathrm{2b}}'}^{(F_{\mathrm{3b}})}  Z_{F_{\mathrm{2b}}',\Sigma_{\mathrm{3b}}}\Bigg)
\end{aligned}
\end{equation}
and
\begin{equation}
\begin{aligned}
C_{F_{\mathrm{3b}}} =& \frac{1}{6 \pi^4} \frac{1}{m \hbar^4} \sum_{F_{\mathrm{2b}}} \left(Z_{F_{\mathrm{2b}},\Sigma_{\mathrm{3b}}}\right)^* a_{F_{\mathrm{2b}}}
\sum_{F_{\mathrm{2b}}'''} W_{F_{\mathrm{2b}},F_{\mathrm{2b}}'''}^{(F_{\mathrm{3b}})} 
\\
&\Bigg( a_{F_{\mathrm{2b}}'''} 
\sum_{F_{\mathrm{2b}}'} W_{F_{\mathrm{2b}}''',F_{\mathrm{2b}}'}^{(F_{\mathrm{3b}})} a_{F_{\mathrm{2b}}'}
 \sum_{F_{\mathrm{2b}}''} a_{F_{\mathrm{2b}}''} W_{F_{\mathrm{2b}}', F_{\mathrm{2b}}''}^{(F_{\mathrm{3b}})}  Z_{F_{\mathrm{2b}}'',\Sigma_{\mathrm{3b}}}
\\
& -\frac{3\sqrt{3}}{2 \pi} (a_{F_{\mathrm{2b}}}''')^2 \sum_{F_{\mathrm{2b}}'} a_{F_{\mathrm{2b}}'}
W_{F_{\mathrm{2b}}''', F_{\mathrm{2b}}'}^{(F_{\mathrm{3b}})}  Z_{F_{\mathrm{2b}}',\Sigma_{\mathrm{3b}}}
 \Bigg).
\end{aligned}
\end{equation}
The relevant three-body spin state $\lvert \Sigma_{\mathrm{3b}} \rangle$ is either $\lvert 1,M_{F_{\mathrm{3b}}} [+] \rangle$ or $\lvert 3 ,M_{F_{\mathrm{3b}}} (2) \rangle$. For $\lvert \Sigma_{\mathrm{3b}} \rangle = \lvert 1,M_{F_{\mathrm{3b}}} [+] \rangle$ we have
\begin{equation}\label{eq:Zvector_F3b=1_full}
\left[Z_{F_{\mathrm{2b}},\Sigma_{\mathrm{3b}}}\right] = \begin{pmatrix}
\frac{\sqrt{5}}{3}\\
0 \\
\frac{2}{3}
\end{pmatrix}
\end{equation}
and
\begin{equation}\label{eq:tildeW_matrix_F3b=1_full}
\left[W_{F_{\mathrm{2b}},F_{\mathrm{2b}}'}^{(1)}\right] = \begin{pmatrix}
\frac{2}{3} & -\frac{2}{3}\sqrt{3} & \frac{2}{3}\sqrt{5} \\
\frac{2}{3}\sqrt{3} & -1 & -\sqrt{\frac{5}{3}} \\
\frac{2}{3}\sqrt{5} & \sqrt{\frac{5}{3}} & \frac{1}{3}
\end{pmatrix},
\end{equation}
where $F_{\mathrm{2b}} = 0, 1, 2$ labels the rows and $F_{\mathrm{2b}}' = 0, 1, 2$ labels the columns. Equation~\eqref{eq:Zvector_F3b=1_full} follows from Eqs.~\eqref{eq:def_Z_F2b_Phi3b} and \eqref{eq:+_to_0_2_F3b=1}. Equation~\eqref{eq:tildeW_matrix_F3b=1_full} can be derived from Eq.~\eqref{eq:def_W_F2b_F2bprime^(F3b)}. A more detailed definition of the permutation operator $P_+^{\mathrm{s}}$ can be found in Appendix~\ref{sec:D1_resonant_p_wave}.

Similarly for $\lvert \Sigma_{\mathrm{3b}} \rangle = \lvert 3,M_{F_{\mathrm{3b}}} (2) \rangle$ we have
\begin{equation}\label{eq:Zvector_F3b=3_full}
\left[Z_{F_{\mathrm{2b}},\Sigma_{\mathrm{3b}}}\right] = \begin{pmatrix}
0\\
0 \\
1
\end{pmatrix}
\end{equation}
and
\begin{equation}\label{eq:tildeW_matrix_F3b=3_full}
\left[W_{F_{\mathrm{2b}},F_{\mathrm{2b}}'}^{(3)}\right] = \begin{pmatrix}
0 & 0 & 0 \\
0 & 0 & 0 \\
0 & 0 & 2
\end{pmatrix}.
\end{equation}
Clearly, $A_{F_{\mathrm{3b}}}$, $B_{F_{\mathrm{3b}}}$ and $C_{F_{\mathrm{3b}}}$ depend only on $a_0$ and $a_2$ for $F_{\mathrm{3b}} = 1$ and only on $a_2$ for $F_{\mathrm{3b}} = 3$.

\section{Integral equations for three identical spin-1 bosons}
\label{sec:integral_eq_spinor}

For three identical spin-1 bosons scattering at zero energy, the relevant transition amplitude is ${}_{\alpha}\langle \mathbf{p}, \mathbf{q}, \Sigma_{\mathrm{3b}} | \breve{U}_{\alpha 0}(0) | \mathbf{0}, \mathbf{0}, \Sigma_{\mathrm{3b}} \rangle$. We expand this amplitude as
\begin{equation}\label{eq:breve_Ualpha0_vs_breve_A_nl_spinor}
\begin{aligned}
{}_{\alpha}\langle \mathbf{p}, &\mathbf{q}, \Sigma_{\mathrm{3b}} | \breve{U}_{\alpha 0}(0) | \mathbf{0}, \mathbf{0}, \Sigma_{\mathrm{3b}} \rangle
\\
&= 3 \sum_{F_{\beta \gamma}} \left(Z_{F_{\beta \gamma},\Sigma_{\mathrm{3b}}}\right)^* \Bigg\{
\langle \mathbf{p} | t_{F_{\beta \gamma}}(0) | \mathbf{0} \rangle \delta(\mathbf{q}) \, Z_{F_{\beta \gamma},\Sigma_{\mathrm{3b}}} 
\\
&+ \sum_{l = 0}^{\infty} \frac{(-1)^l}{\sqrt{2 l + 1}}  
\sum_{m_l  = -l}^{l} 4 \pi \,Y_l^{m_l}(\hat{\mathbf{p}}) \left[Y_{l}^{m_l}(\hat{\mathbf{q}})\right]^*
\\
&\sum_{n = 1}^{\infty} \tau_{n l,F_{\beta \gamma}}\left(-\frac{3 q^2}{4 m}\right) g_{n l, F_{\beta \gamma}}\left(p, -\frac{3 q^2}{4 m}\right) \breve{A}_{n l,F_{\beta \gamma}}^{(F_{\mathrm{3b}})}(q) \Bigg\},
\end{aligned}
\end{equation}
where $\tau_{n l,F_{\beta \gamma}}\left(z_{\mathrm{2b}}\right)$ and $g_{n l, F_{\beta \gamma}}\left(p, z_{\mathrm{2b}}\right)$ are defined by expanding $t_{F_{\mathrm{2b}}}\left(z_{\mathrm{2b}}\right)$ as
\begin{equation}\label{eq:expansion_t_Fbetagamma}
\begin{aligned}
t_{F_{\mathrm{2b}}}(z_{\mathrm{2b}}) &= -4 \pi \sum_{l=0}^{\infty} \sum_{n = 1}^{\infty} \tau_{n l, F_{\mathrm{2b}}}\left(z_{\mathrm{2b}}\right) 
\\
&\sum_{m_l = -l}^{l}  \lvert g_{n l m_l, F_{\mathrm{2b}}}\left(z_{\mathrm{2b}}\right) \rangle \langle g_{n l m_l, F_{\mathrm{2b}}}\left(z_{\mathrm{2b}}\right) \rvert
\end{aligned}
\end{equation}
with $\langle \mathbf{p} | g_{n l m_l, F_{\mathrm{2b}}}\left(z_{\mathrm{2b}}\right) \rangle = Y_l^{m_l}(\hat{\mathbf{p}}) g_{n l, F_{\mathrm{2b}}}\left(p,z_{\mathrm{2b}}\right)$. This expansion can be done in various ways \cite{mestrom2019squarewell} and we take the Weinberg expansion \cite{weinberg1963expansion,mestrom2019squarewell}.

The functions $\breve{A}_{n l,F_{\beta \gamma}}^{(F_{\mathrm{3b}})}(q)$ can be determined from Eqs.~\eqref{eq:Ubreve_BBB_FFF} and \eqref{eq:breve_Ualpha0_vs_breve_A_nl_spinor}. This results in the integral equation
\begin{equation}\label{eq:breve_mathsfA_nl_q_spinor}
\begin{aligned}
\breve{A}_{n l,F_{\beta \gamma}}^{(F_{\mathrm{3b}})}&(q) = - \Delta_{l + F_{\beta \gamma}} \sum_{n' F_{\beta \gamma}'} \tau_{n',0,F_{\beta \gamma}'}(0) g_{n',0,F_{\beta \gamma}'}(0,0)
\\
& U_{n l, n', 0}^{(F_{\beta \gamma},F_{\beta \gamma}')}(q,0) W_{F_{\beta \gamma},F_{\beta \gamma}'}^{(F_{\mathrm{3b}})}   Z_{F_{\beta \gamma}',\Sigma_{\mathrm{3b}}}
\\
&+ 4 \pi \Delta_{l + F_{\beta \gamma}} \sum_{n' l' F_{\beta \gamma}'} \int_0^{\infty} \tau_{n' l',F_{\beta \gamma}'}\left(-\frac{3 q'^2}{4 m}+ i 0\right)
\\
& U_{n l, n' l'}^{(F_{\beta \gamma},F_{\beta \gamma}')}(q,q') W_{F_{\beta \gamma},F_{\beta \gamma}'}^{(F_{\mathrm{3b}})} (-1)^{l'}  \breve{A}_{n' l', F_{\beta \gamma}'}^{(F_{\mathrm{3b}})}(q') \,q'^2 \,dq',
\end{aligned}
\end{equation}
where $\Delta_{l + F_{\beta \gamma}} = \left(1 + (-1)^{l + F_{\beta \gamma}} \right)/2$ ensures that even (odd) $l$ is combined with even (odd) $F_{\mathrm{2b}}$ and where
\begin{equation} \label{eq:U_nln1l1^alpha_alpha_spinor}
\begin{aligned}
U_{n l,n' l'}^{(F_{\beta \gamma},F_{\beta \gamma}')}&(q,q') = \frac{m}{4 \pi} (-1)^{l + l'} \sqrt{2 l + 1} \sqrt{2 l' + 1}
\\
&\int P_{l}(\hat{\mathbf{q}} \cdot \reallywidehat{\mathbf{q}' + \frac{1}{2} \mathbf{q}})
P_{l'}(\reallywidehat{\mathbf{q} + \frac{1}{2} \mathbf{q}'} \cdot \hat{\mathbf{q}}')
\\
&\frac{1}{q^2 + q'^2 + \mathbf{q}\cdot \mathbf{q}'}
g_{n l,F_{\beta \gamma}}\left(|\mathbf{q}' + \frac{1}{2} \mathbf{q}|,-\frac{3 q^2}{4 m}\right)
\\
&g_{n' l',F_{\beta \gamma}'}\left(|\mathbf{q} + \frac{1}{2} \mathbf{q}'|,-\frac{3 q'^2}{4 m}\right)
\,d\hat{\mathbf{q}}'
\end{aligned}
\end{equation}
with the Legendre polynomials $P_l(x)$. This integral equation is solved as a matrix equation by discretizing the momentum $q$. The scattering hypervolume $D_{F_{\mathrm{3b}}}$ can be extracted from the solution using the same approach as presented in Ref.~\cite{mestrom2019hypervolumeSqW} for three identical spinless bosons.

\section{Connection to the BBX system}
\label{sec:spinor_connection_BBX}

The three-body integral equations for identical spin-1 bosons with $F_{\mathrm{3b}} = 1$ reduce to those corresponding to spinless bosons when $V_{F_{\mathrm{2b}} = 0} = V_{F_{\mathrm{2b}} = 2}$. However, these integral equations are also very closely related to those corresponding to two identical spinless bosons (B) and one dissimilar spinless particle (X) which we indicate as the BBX system. This can be seen as follows. The spinor system has three relevant interactions, namely the even partial-wave components of $V_{F_{\mathrm{2b}} = 0}$ and $V_{F_{\mathrm{2b}} = 2}$ and the odd partial-wave components of $V_{F_{\mathrm{2b}} = 1}$. Similarly, the scattering properties of the BBX system depend on the even partial-wave components of the BB interaction $V_{\mathrm{BB}}$ and the BX interaction $V_{\mathrm{BX}}$ as well as on the odd partial-wave components of $V_{\mathrm{BX}}$. This suggests that we can map the three-body scattering problem for identical spin-1 bosons onto the one for the BBX system. 

The three-body integral equations for the BBX system with equal masses (i.e., $m_{\mathrm{X}} = m_{\mathrm{B}}$) are also given by Eq.~\eqref{eq:breve_mathsfA_nl_q_spinor}, but now $F_{\mathrm{2b}}$ does not label the two-body spin. Instead, $F_{\mathrm{2b}}$ labels the two-body configuration which is either BX ($F_{\mathrm{2b}} = 0$ and 1) or BB ($F_{\mathrm{2b}} = 2$). Furthermore, this mapping requires
\begin{equation}\label{eq:Zvector_BBX_full}
\left[Z_{F_{\mathrm{2b}},\Sigma_{\mathrm{3b}}}\right] = \begin{pmatrix}
1\\
0 \\
1
\end{pmatrix}
\end{equation}
and
\begin{equation}\label{eq:tildeW_matrix_BBX_full}
\left[W_{F_{\mathrm{2b}},F_{\mathrm{2b}}'}^{(F_{\mathrm{3b}})}\right] = \begin{pmatrix}
1 & -1 & 1 \\
1 & -1 & -1 \\
2 & 2 & 0
\end{pmatrix}.
\end{equation}
This connection can be derived from Ref.~\cite{mestrom2021hypervolumeBBX}, in which the three-body transition amplitude of the BBX system was analyzed. 

Although the zero-energy three-body transition amplitude of the BBX system with $m_{\mathrm{X}} = m_{\mathrm{B}}$ is determined by Eq.~\eqref{eq:breve_mathsfA_nl_q_spinor}, the behavior of the corresponding three-body scattering hypervolume can be very different from $D_1$ corresponding to the spinor system with $F_{\mathrm{3b}} = 1$ due to the difference in $W_{F_{\mathrm{2b}},F_{\mathrm{2b}}'}^{(F_{\mathrm{3b}})}$. For example, the Efimov effect occurs in the BBX system whenever the scattering length $a_{\mathrm{BX}}$ corresponding to $V_{\mathrm{BX}}$ diverges \cite{mestrom2021hypervolumeBBX}, whereas the Efimov effect in the spinor system with $F_{\mathrm{3b}} = 1$ only occurs when both $a_0$ and $a_2$ diverge simultaneously \cite{colussi2014EfimovSpinor,colussi2016EfimovSpinor}. This difference can also be understood from the simple toy model presented in Ref.~\cite{colussi2016toymodelspinor}.

We note that our definition of the three-body scattering hypervolume in Eq.~\eqref{eq:def_D_F3b_spinor} can be generalized to the BBX system with unequal masses (i.e., $m_{\mathrm{X}} \neq m_{\mathrm{B}}$) as is done in Ref.~\cite{tan2021hypervolumeBBX}. This scattering hypervolume can be important for determining the stability of ultracold mixtures against collapse or phase separation \cite{tan2021hypervolumeBBX}.

\section{Special cases}
\label{sec:special_cases}

Here we determine the behavior of $D_1$ in the special cases that $V_{F_{\mathrm{2b}} = 0} = V_{F_{\mathrm{2b}} = 2}$ or $a_0 = a_2$. First of all, we note that $D_1 = D_3$ for $V_{F_{\mathrm{2b}} = 0} = V_{F_{\mathrm{2b}} = 2}$ when taking $\rho_1 = \rho_3$. This is evident from
\begin{widetext}
\begin{eqnarray}\label{eq:Talpha_+-_states}
\begin{aligned}
{}_{\alpha}\langle \mathbf{p}, \mathbf{q}, 1, M_{F_{\mathrm{3b}}} [+] | T_{\alpha}(z) | \mathbf{p}', \mathbf{q}', 1, M_{F_{\mathrm{3b}}} [+] \rangle_{\alpha} 
&= \langle \mathbf{q} | \mathbf{q}'\rangle \Bigg(\frac{5}{9} \langle \mathbf{p} | t_{0}\left(z-\frac{3 q^2}{4 m}\right) | \mathbf{p}' \rangle + \frac{4}{9} \langle \mathbf{p} | t_{2}\left(z-\frac{3 q^2}{4 m}\right) | \mathbf{p}' \rangle \Bigg)
\\
{}_{\alpha}\langle \mathbf{p}, \mathbf{q}, 1, M_{F_{\mathrm{3b}}} [+] | T_{\alpha}(z) | \mathbf{p}', \mathbf{q}', 1, M_{F_{\mathrm{3b}}} [-] \rangle_{\alpha} 
&= \langle \mathbf{q} | \mathbf{q}'\rangle \frac{2 \sqrt{5}}{9}\Bigg(\langle \mathbf{p} | t_{0}\left(z-\frac{3 q^2}{4 m}\right) | \mathbf{p}' \rangle - \langle \mathbf{p} | t_{2}\left(z-\frac{3 q^2}{4 m}\right) | \mathbf{p}' \rangle \Bigg)
\\
{}_{\alpha}\langle \mathbf{p}, \mathbf{q}, 1, M_{F_{\mathrm{3b}}} [-] | T_{\alpha}(z) | \mathbf{p}', \mathbf{q}', 1, M_{F_{\mathrm{3b}}} [+] \rangle_{\alpha} 
&= \langle \mathbf{q} | \mathbf{q}'\rangle \frac{2 \sqrt{5}}{9}\Bigg(\langle \mathbf{p} | t_{0}\left(z-\frac{3 q^2}{4 m}\right) | \mathbf{p}' \rangle - \langle \mathbf{p} | t_{2}\left(z-\frac{3 q^2}{4 m}\right) | \mathbf{p}' \rangle \Bigg)
\\
{}_{\alpha}\langle \mathbf{p}, \mathbf{q}, 1, M_{F_{\mathrm{3b}}} [-] | T_{\alpha}(z) | \mathbf{p}', \mathbf{q}', 1, M_{F_{\mathrm{3b}}} [-] \rangle_{\alpha} 
&= \langle \mathbf{q} | \mathbf{q}'\rangle \Bigg(\frac{4}{9} \langle \mathbf{p} | t_{0}\left(z-\frac{3 q^2}{4 m}\right) | \mathbf{p}' \rangle + \frac{5}{9} \langle \mathbf{p} | t_{2}\left(z-\frac{3 q^2}{4 m}\right) | \mathbf{p}' \rangle \Bigg),
\end{aligned}
\end{eqnarray}
\end{widetext}
which shows that $T_{\alpha}(z)$ is diagonal in the spin states $\lvert 1, M_{F_{\mathrm{3b}}} [\pm]\rangle$ for $V_{F_{\mathrm{2b}} = 0} = V_{F_{\mathrm{2b}} = 2}$. Since $P$ and $G_0(z)$ are also diagonal in the spin state $\lvert 1, M_{F_{\mathrm{3b}}} [+]\rangle$, $D_1$ connects directly to the three-body scattering hypervolume for spinless bosons interacting via pairwise potentials $V_{F_{\mathrm{2b}} = 0} = V_{F_{\mathrm{2b}} = 2}$, so that $D_1 = D_3$ for $\rho_1 = \rho_3$.

Secondly, we consider the case that $a_0 = a_2$ and take $\rho_1 = \rho_3 = |a_2|$. This condition does not necessarily mean that $V_{F_{\mathrm{2b}} = 0} = V_{F_{\mathrm{2b}} = 2}$, so that $D_1 \neq D_3$ in general. However, in the limit $|a_0| \to \infty$ the parts of $D_1$ and $D_3$ that are purely determined by $a_0$ and $a_2$ are still connected. These parts are indicated by $D_{1,\mathrm{hh}}$ and $D_{3,\mathrm{hh}}$ because they originate from hard-hyperspherelike collisions \cite{dincao2018review}. From Refs.~\cite{mestrom2019hypervolumeSqW,mestrom2020hypervolumeVdW} it follows that $D_{3,\mathrm{hh}} = 1689a_2^4$ for $|a_2| \to \infty$. To determine $D_{1,\mathrm{hh}}$, we analyze Eq.~\eqref{eq:Talpha_+-_states} for the interaction $V_{F_{\mathrm{2b}}}$ in Eq.~\eqref{eq:V_sepCut_F2b}.

From the Lippmann-Schwinger equation in Eq.~\eqref{eq:LS_tF2b} with $V_{F_{\mathrm{2b}}}$ in Eq.~\eqref{eq:V_sepCut_F2b}, we find that
\begin{equation}
t_{F_{\mathrm{2b}}}(z_{\mathrm{2b}}) = -\tau_{F_{\mathrm{2b}}}(z_{\mathrm{2b}}) \lvert g \rangle \langle g \lvert,
\end{equation}
where
\begin{equation}
\tau_{F_{\mathrm{2b}}}(z_{\mathrm{2b}}) = -\frac{1}{4 \pi^2 \mu \hbar} \frac{a_{F_{\mathrm{2b}}}}{1 - \frac{2}{\pi}a_{F_{\mathrm{2b}}} q_{\mathrm{2b}}/\hbar \arctan\left(\Lambda/q_{\mathrm{2b}}\right)}
\end{equation}
with $q_{\mathrm{2b}} \equiv - i \sqrt{2 \mu z_{\mathrm{2b}}}$, $\mu = m/2$ is the two-particle reduced mass and
\begin{equation}
a_{F_{\mathrm{2b}}} = -\frac{4 \pi^2 \mu \hbar \zeta_{F_{\mathrm{2b}}}}{1 - 8 \pi \mu \Lambda \zeta_{F_{\mathrm{2b}}}}.
\end{equation}
Since
\begin{equation}
\begin{aligned}
\frac{5}{9} \tau_{0}(z_{\mathrm{2b}}) + \frac{4}{9}\tau_{2}(z_{\mathrm{2b}}) =& -\frac{1}{4 \pi^2 \mu \hbar} \frac{\tilde{a}}{1 - \frac{2}{\pi} \frac{\tilde{a} q_{\mathrm{2b}}}{\hbar} \arctan\left(\Lambda/q_{\mathrm{2b}}\right)} 
\\
&+ O\left(X^2\right),
\end{aligned}
\end{equation}
\begin{equation}
\begin{aligned}
\frac{4}{9} \tau_{0}(z_{\mathrm{2b}}) + \frac{5}{9}\tau_{2}(z_{\mathrm{2b}}) =& -\frac{1}{4 \pi^2 \mu \hbar} \frac{\tilde{a}}{1 - \frac{2}{\pi} \frac{\tilde{a} q_{\mathrm{2b}}}{\hbar} \arctan\left(\Lambda/q_{\mathrm{2b}}\right)} 
\\
&+ O\left(X\right)
\end{aligned}
\end{equation}
and $\tau_{0}(z_{\mathrm{2b}}) - \tau_{2}(z_{\mathrm{2b}}) = O\left(X\right)$, it follows from Eqs.~\eqref{eq:Ubreve_BBB_FFF} and \eqref{eq:Talpha_+-_states} that the transition amplitude ${}_{\alpha}\langle \mathbf{p}, \mathbf{q}, 1, M_{F_{\mathrm{3b}}} [+] | \breve{U}_{\alpha 0}(0) | \mathbf{0}, \mathbf{0}, 1, M_{F_{\mathrm{3b}}} [+] \rangle$ maps onto ${}_{\alpha}\langle \mathbf{p}, \mathbf{q} | \breve{U}_{\alpha 0}(0) | \mathbf{0}, \mathbf{0} \rangle$ corresponding to spinless bosons not only for $X = 0$, but also up to the first order in $X$. This results in Eq.~\eqref{eq:Dh_hh_F3b_1_Taylor_X}.

\section{$D_1$ for resonant $p$-wave interactions}
\label{sec:D1_resonant_p_wave}

In this appendix we follow the approach of Ref.~\cite{mestrom2021hypervolumeBBX} to determine the behavior of $D_1$ near a $p$-wave dimer resonance with $F_{\mathrm{2b}} = 1$. Its dominant behavior scales as $\sqrt{-v_1}$ which is solely determined by scattering events described by $T_{\alpha}(0) G_0(0) P T_{\alpha}(0) G_0(0) P T_{\alpha}(0)$ for $\alpha = 1$, 2 and 3 where the middle $T_{\alpha}$ corresponds to $F_{\mathrm{2b}} = 1$. After all, the first and final $T_{\alpha}$ can only contribute to $D_1$ via their $s$-wave components. Scattering processes described by other diagrams do not result in a $\sqrt{-v_1}$ scaling of $D_1$ near a $p$-wave dimer resonance in a similar way as was found for dissimilar particles with resonant $p$-wave interactions in Ref.~\cite{mestrom2021hypervolumeBBX}. So to study the effect of resonant $p$-wave interactions on $D_1$, we analyze 
\begin{equation}\label{eq:TG0PTG0PT_v1}
\begin{aligned}
\mathcal{T}_{\alpha} &\equiv \langle \mathbf{0},\mathbf{0}, 1, M_{F_{\mathrm{3b}}} [+] | T_{\alpha}(0) G_0(0) P 
\Bigg( \lvert  1, M_{F_{\mathrm{3b}}} (1) \rangle_{\alpha} 
\\
&\, {}_{\alpha}\langle 1, M_{F_{\mathrm{3b}}} (1) | 
T_{\alpha}(0) |  1, M_{F_{\mathrm{3b}}} (1) \rangle_{\alpha} 
\, {}_{\alpha}\langle 1, M_{F_{\mathrm{3b}}} (1) \rvert
\Bigg) 
\\
&G_0(0) P T_{\alpha}(0) | \mathbf{0}, \mathbf{0}, 1, M_{F_{\mathrm{3b}}} [+] \rangle,
\end{aligned}
\end{equation}
where we used $\mathcal{T}_{\alpha}$ as shorthand notation for the matrix element of interest. Considering the spin basis $\big\{\lvert 1, M_{F_{\mathrm{3b}}} [+] \rangle_{\alpha}$, $\lvert 1, M_{F_{\mathrm{3b}}} [-] \rangle_{\alpha}$, $\lvert 1, M_{F_{\mathrm{3b}}} (1) \rangle_{\alpha}\big\}$, $G_0(z)$ is diagonal in all three-body spin states, but $T_{\alpha}(z)$ is only diagonal in $\lvert 1, M_{F_{\mathrm{3b}}} (1) \rangle_{\alpha}$ and $P$ is only diagonal in the fully symmetric spin states $\lvert 1, M_{F_{\mathrm{3b}}} [+] \rangle$. Therefore,
\begin{equation}\label{eq:TG0PTG0PT_v2}
\begin{aligned}
\mathcal{T}_{\alpha} =& \langle \mathbf{0},\mathbf{0}, 1, M_{F_{\mathrm{3b}}} [+] | T_{\alpha}(0) | 1, M_{F_{\mathrm{3b}}} [-] \rangle_{\alpha} \,
\\
&
G_0^{\mathrm{c}}(0) \, {}_{\alpha}\langle 1, M_{F_{\mathrm{3b}}} [-] | P |  1, M_{F_{\mathrm{3b}}} (1) \rangle_{\alpha}
\\
&{}_{\alpha}\langle 1, M_{F_{\mathrm{3b}}} (1) | 
T_{\alpha}(0) |  1, M_{F_{\mathrm{3b}}} (1) \rangle_{\alpha} 
\\
& G_0^{\mathrm{c}}(0)
\, {}_{\alpha}\langle 1, M_{F_{\mathrm{3b}}} (1) | P | 1, M_{F_{\mathrm{3b}}} [-] \rangle_{\alpha} 
\\
&
{}_{\alpha}\langle 1, M_{F_{\mathrm{3b}}} [-] | T_{\alpha}(0) | \mathbf{0}, \mathbf{0}, 1, M_{F_{\mathrm{3b}}} [+] \rangle,
\end{aligned}
\end{equation}
where we defined $G_0^{\mathrm{c}}(z)$ as the part of $G_0(z)$ acting on coordinate space.
Defining $P_{\alpha \beta}$ as the permutation operator of particles $\alpha$ and $\beta$, we can write $P = P_+ + P_-$ where $P_+ = P_{\alpha \gamma} P_{\beta \gamma}$ permutates the particles according to $\alpha \beta \gamma \to \gamma \alpha \beta$ and $P_- = P_{\alpha \beta} P_{\beta \gamma}$ according to $\alpha \beta \gamma \to \beta \gamma \alpha$. 
We define $P_{\pm}^{\mathrm{c}}$ and $P_{\pm}^{\mathrm{s}}$ as the parts of $P_{\pm}$ acting on coordinate and spin space, respectively, so that $P_{\pm} = P_{\pm}^{\mathrm{c}} P_{\pm}^{\mathrm{s}}$. Consequently, we have
\begin{equation}\label{eq:P_pm_spin_F3b_1}
\begin{aligned}
{}_{\alpha}\langle 1, M_{F_{\mathrm{3b}}} [-] | P_{\pm}^{\mathrm{s}} | 1, M_{F_{\mathrm{3b}}} (1) \rangle_{\alpha} &= \mp \frac{1}{2}\sqrt{3}
\\
{}_{\alpha}\langle 1, M_{F_{\mathrm{3b}}} (1) | P_{\pm}^{\mathrm{s}} | 1, M_{F_{\mathrm{3b}}} [-] \rangle_{\alpha} &= \pm \frac{1}{2}\sqrt{3}.
\end{aligned}
\end{equation}
Next, we work out Eq.~\eqref{eq:TG0PTG0PT_v2} further, resulting in
\begin{widetext}
\begin{eqnarray}\label{eq:TG0PTG0PT_v3}
\begin{aligned}
\mathcal{T}_{\alpha} =&  \, m^2 \int \frac{1}{q'^4}
\Bigg\{
\langle \mathbf{0} | \tilde{t}(0) | -\mathbf{q}' \rangle 
\, {}_{\alpha}\langle 1, M_{F_{\mathrm{3b}}} [-] | P^{\mathrm{s}}_{-} | 1, M_{F_{\mathrm{3b}}} (1) \rangle_{\alpha}\,
\langle \frac{1}{2}\mathbf{q}' | t_{1}\left(-\frac{3 q'^2}{4 m}\right) | -\frac{1}{2}\mathbf{q}'\rangle
\\
&
{}_{\alpha}\langle 1, M_{F_{\mathrm{3b}}} (1) | P^{\mathrm{s}}_{-} | 1, M_{F_{\mathrm{3b}}} [-] \rangle_{\alpha} \,
\langle \mathbf{q}' | \tilde{t}(0) |\mathbf{0} \rangle
\\
&+
\langle \mathbf{0} | \tilde{t}(0) |\mathbf{q}' \rangle 
\,{}_{\alpha}\langle 1, M_{F_{\mathrm{3b}}} [-] | P^{\mathrm{s}}_{+} | 1, M_{F_{\mathrm{3b}}} (1) \rangle_{\alpha}\,
\langle -\frac{1}{2}\mathbf{q}' | t_{1}\left(-\frac{3 q'^2}{4 m}\right) | -\frac{1}{2}\mathbf{q}'\rangle
\\
&
{}_{\alpha}\langle 1, M_{F_{\mathrm{3b}}} (1) | P^{\mathrm{s}}_{-} | 1, M_{F_{\mathrm{3b}}} [-] \rangle_{\alpha}\,
\langle \mathbf{q}' | \tilde{t}(0) |\mathbf{0} \rangle
\\
&+
\langle \mathbf{0} | \tilde{t}(0) | -\mathbf{q}' \rangle 
\,{}_{\alpha}\langle 1, M_{F_{\mathrm{3b}}} [-] | P^{\mathrm{s}}_{-} | 1, M_{F_{\mathrm{3b}}} (1) \rangle_{\alpha}\,
\langle \frac{1}{2}\mathbf{q}' | t_{1}\left(-\frac{3 q'^2}{4 m}\right) | \frac{1}{2}\mathbf{q}'\rangle
\\
&
{}_{\alpha}\langle 1, M_{F_{\mathrm{3b}}} (1) | P^{\mathrm{s}}_{+} | 1, M_{F_{\mathrm{3b}}} [-] \rangle_{\alpha}\,
\langle -\mathbf{q}' | \tilde{t}(0) |\mathbf{0} \rangle
\\
&+
\langle \mathbf{0} | \tilde{t}(0) |\mathbf{q}' \rangle 
\,{}_{\alpha}\langle 1, M_{F_{\mathrm{3b}}} [-] | P^{\mathrm{s}}_{+} | 1, M_{F_{\mathrm{3b}}} (1) \rangle_{\alpha}\,
\langle -\frac{1}{2}\mathbf{q}' | t_{1}\left(-\frac{3 q'^2}{4 m}\right) | \frac{1}{2}\mathbf{q}'\rangle
\\
&
{}_{\alpha}\langle 1, M_{F_{\mathrm{3b}}} (1) | P^{\mathrm{s}}_{+} | 1, M_{F_{\mathrm{3b}}} [-] \rangle_{\alpha}\,
\langle -\mathbf{q}' | \tilde{t}(0) |\mathbf{0} \rangle
\Bigg\}\,d\mathbf{q}',
\end{aligned}
\end{eqnarray}
\end{widetext}
where we used Eq.~\eqref{eq:Talpha_+-_states} and defined
\begin{equation}\label{eq:def_tilde_t}
\begin{aligned}
\langle \mathbf{p} | \tilde{t}(z_{\mathrm{2b}}) |\mathbf{p}' \rangle &= \frac{2\sqrt{5}}{9} \Bigg(\langle \mathbf{p} | t_{0}(z_{\mathrm{2b}}) |\mathbf{p}' \rangle - \langle \mathbf{p} | t_{2}(z_{\mathrm{2b}}) |\mathbf{p}' \rangle\Bigg)
\end{aligned}
\end{equation}
for notational convenience. We consider spherically symmetric potentials for which
\begin{equation}
\langle \mathbf{p} | t_{F_{\mathrm{2b}}}(z_{\mathrm{2b}}) | \mathbf{p}'\rangle = \sum_{l = 0}^{\infty} (2 l + 1) P_l(\hat{\mathbf{p}}\cdot \hat{\mathbf{p}}') t_{l,F_{\mathrm{2b}}}\left(p,p',z_{\mathrm{2b}}\right),
\end{equation}
\begin{equation}
t_{l,F_{\mathrm{2b}}}\left(p,p',z_{\mathrm{2b}}\right) = t_{l,F_{\mathrm{2b}}}\left(p',p,z_{\mathrm{2b}}\right),
\end{equation}
\begin{equation}\label{eq:tlneq0_zero_p}
t_{l\neq 0,F_{\mathrm{2b}}}\left(0,p',z_{\mathrm{2b}}\right) = 0,
\end{equation}
\begin{equation}\label{eq:t0_small_p_pa_z}
\begin{aligned}
 t_{0,F_{\mathrm{2b}}}&\left(p,p',-\frac{\hbar^2 \kappa^2}{m}\right) = \frac{\frac{a_{F_{\mathrm{2b}}}}{2 \pi^2 m \hbar} + O\left(p^2, (p')^2\right)}{1 - a_{F_{\mathrm{2b}}} \kappa + O\left(\kappa^2\right)}
\end{aligned}
\end{equation}
and
\begin{equation}\label{eq:t1_small_p_pa_z}
\begin{aligned}
 t_{1,F_{\mathrm{2b}}}&\left(p,p',-\frac{\hbar^2 \kappa^2}{m}\right) = \left(\frac{v_{F_{\mathrm{2b}}} p p'}{2 \pi^2 m \hbar^3} + O\left(p (p')^3, p' p^3\right) \right)
\\
&
\frac{1}{1 - \frac{1}{2} \tilde{r}_{F_{\mathrm{2b}}} v_{F_{\mathrm{2b}}} \kappa^2 + v_{F_{\mathrm{2b}}} \kappa^3 + O\left(\kappa^4\right)}.
\end{aligned}
\end{equation}
These conditions apply in general to short-range potentials \cite{taylor1972scattering}. Eq.~\eqref{eq:TG0PTG0PT_v3} now simplifies to
\begin{equation}\label{eq:TG0PTG0PT_v4}
\begin{aligned}
\mathcal{T}_{\alpha} =&  \,3 m^2 \int \frac{1}{q'^4}
\left|\langle \mathbf{0} | \tilde{t}(0) | \mathbf{q}' \rangle\right|^2
\\
&\langle \frac{1}{2}\mathbf{q}' | t_{1}\left(-\frac{3 q'^2}{4 m}\right) | \frac{1}{2}\mathbf{q}'\rangle
\,d\mathbf{q}',
\end{aligned}
\end{equation}
where we also used Eq.~\eqref{eq:P_pm_spin_F3b_1}. To get the largest scaling in $v_1$, we use the special property \cite{mestrom2021hypervolumeBBX} that
\begin{align}
\Bigg(\int_0^{Q} &\frac{1}{q^2}  t_{1,F_{\mathrm{2b}}}\left(\frac{1}{2} q,\frac{1}{2} q,-\frac{3 q^2}{4 m} + i 0\right) \,dq \Bigg)\frac{1}{\sqrt{-v_{F_{\mathrm{2b}}}}} \nonumber
\\
&\underset{v_{F_{\mathrm{2b}}} \to \pm\infty}{=} - \frac{\sqrt{6}}{24 \pi} \frac{1}{m \hbar^2 \sqrt{\tilde{r}_{F_{\mathrm{2b}}}}} \label{eq:int_1/q^2_t1_dq_a1abs_to_inf_F2b}
\end{align}
for any positive upper limit $Q$. This behavior originates from an arbitrarily small integration interval, so that $Q$ can be chosen arbitrarily small. Therefore we derive from Eq.~\eqref{eq:TG0PTG0PT_v4} that
\begin{align}
\mathcal{T}_{\alpha}/\sqrt{-v_1} &\underset{v_{1} \to \pm\infty}{=}  \,36 \pi m^2 \left|\langle \mathbf{0} | \tilde{t}(0) | \mathbf{0} \rangle\right|^2 
\left(-\frac{\sqrt{6}}{24 \pi} \frac{1}{m \hbar^2 \sqrt{\tilde{r}_1}}\right) \nonumber
\\
&= -\frac{5 \sqrt{6}}{54 \pi^4} \frac{1}{m \hbar^4 \sqrt{\tilde{r}_1}}\left(a_0 - a_2 \right)^2. \label{eq:TG0PTG0PT_pwave_resonance}
\end{align}
This result together with the definition of $D_1$ in Eq.~\eqref{eq:def_D_F3b_spinor} gives Eq.~\eqref{eq:D1_limit_large_a1}. We note that the limit $v_{1} \to +\infty$ in Eq.~\eqref{eq:TG0PTG0PT_pwave_resonance} can also be derived from the optical theorem for three-particle scattering in a similar way as was done in Ref.~\cite{mestrom2021hypervolumeBBX}, giving another proof that the dominant decay process is three-body recombination into the shallow $p$-wave dimer state.

\section{Additional results for $D_1$}
\label{sec:SqW_D1_R0_neq_R2}

\begin{figure*}[hbtp]
	\begin{subfigure}
	\centering
	\includegraphics[width=3.4in]{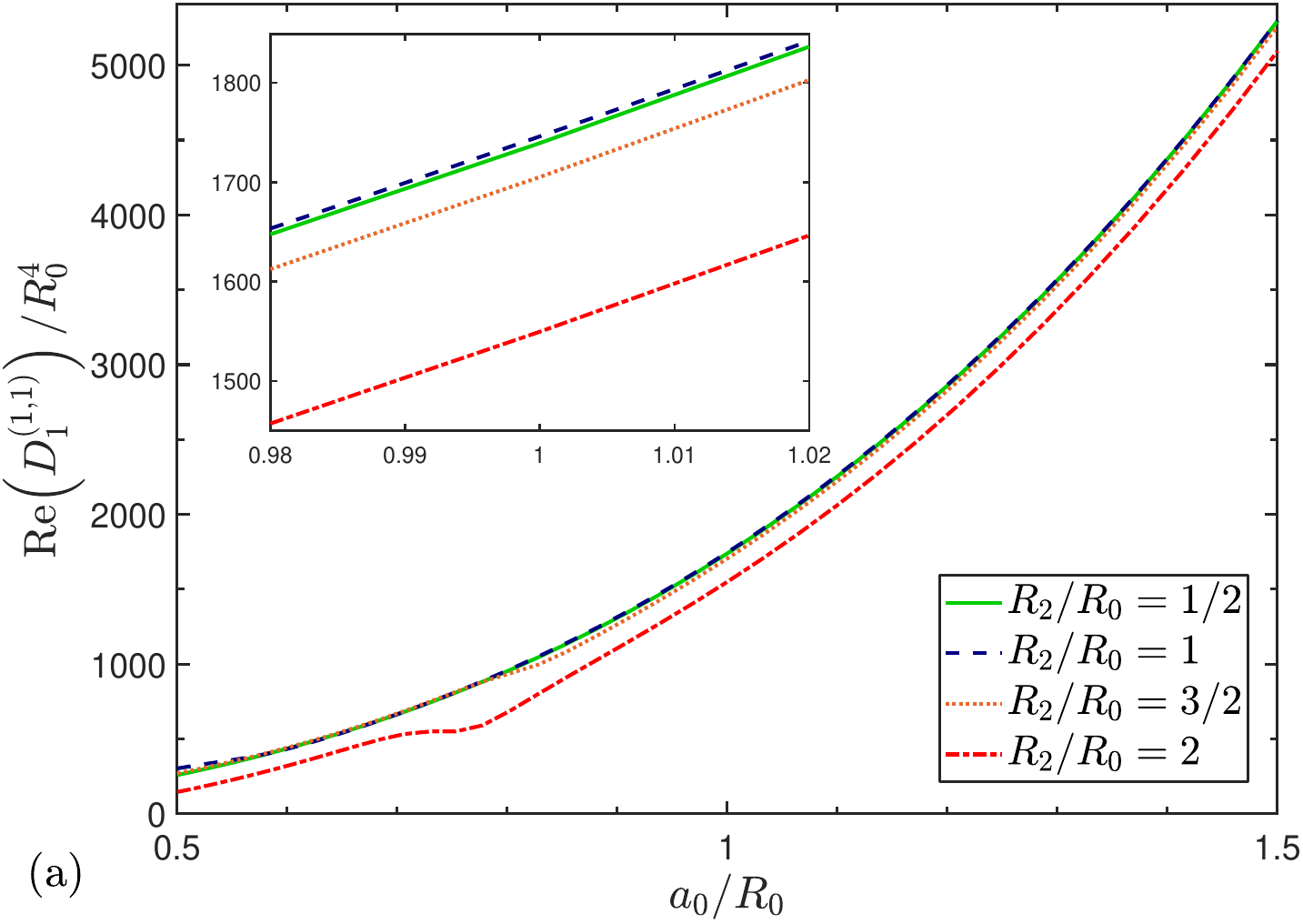}
	\end{subfigure}
\quad
	\begin{subfigure}
	\centering
	\includegraphics[width=3.4in]{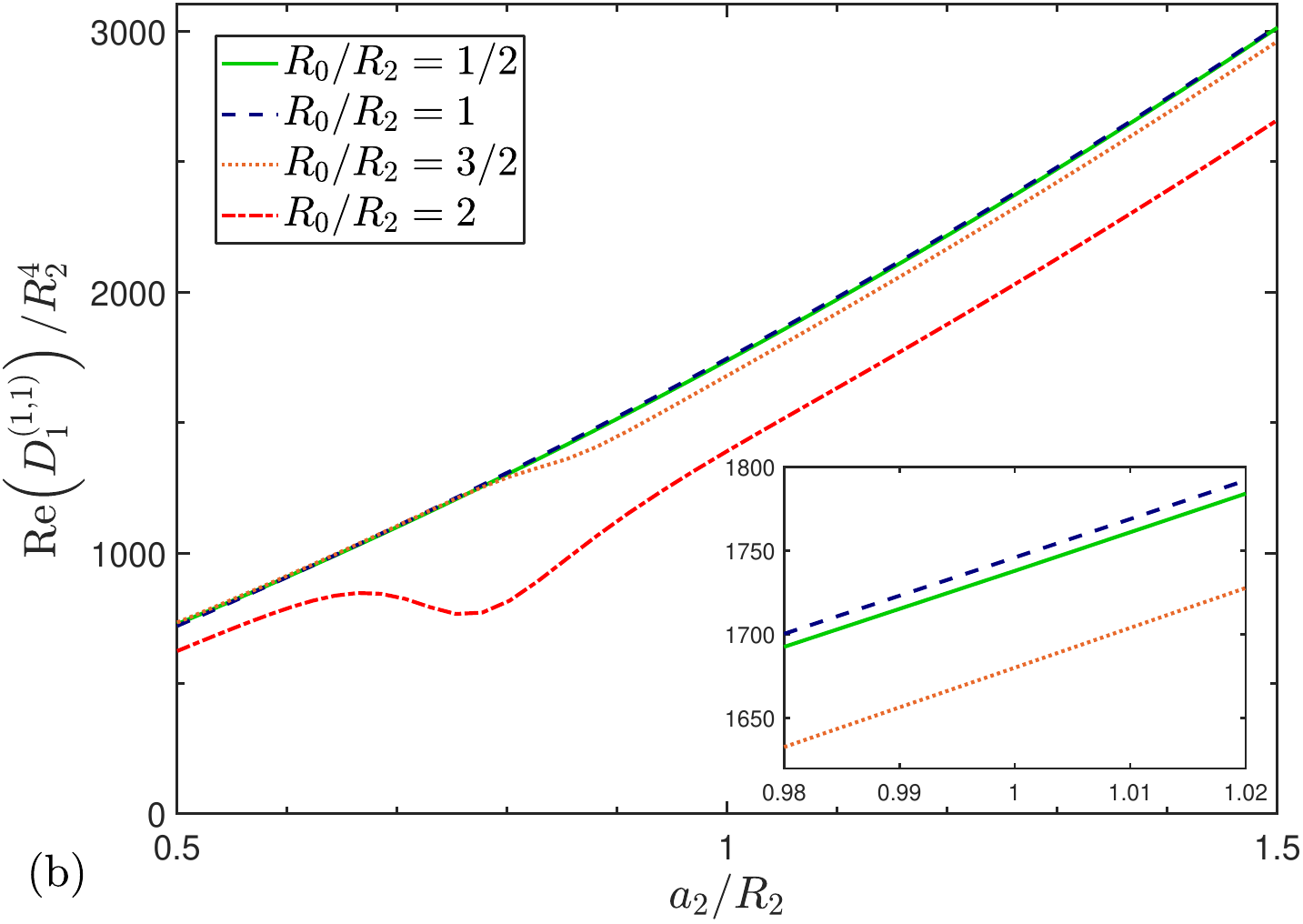}
	\end{subfigure}
    \caption{Real part of the three-body scattering hypervolume $D_1^{(1,1)}$ corresponding to pairwise square-well potentials with various $R_0$ and $R_2$ for (a) $1/2 \leq a_0/R_0 \leq 3/2$ and $a_2/R_0 = 1$ and (b) $1/2 \leq a_2/R_2 \leq 3/2$ and $a_0/R_2 = 1$. We set $V_{F_{\mathrm{2b}} = 1} = 0$ and $\rho_1 = |a_2|$.}
    \label{fig:Dhyp_spinor_SqW_R0neqR2}
\end{figure*}

In Section~\ref{ssec:results_weak_interactions} we have analyzed $\mathrm{Re}\left(D_1\right)$ for potentials $V_{F_{\mathrm{2b}}}$ with the same interaction range. Here we consider $V_{F_{\mathrm{2b}}}$ in Eq.~\eqref{eq:V_F2b_SqW} with $R_0 \neq R_2$. Our results for $\mathrm{Re}\left(D_1\right)$ are presented in Fig.~\ref{fig:Dhyp_spinor_SqW_R0neqR2}. This figure shows that the curves for $\mathrm{Re}\left(D_1\right)$ shift when $R_0$ or $R_2$ becomes larger than $a_0$ or $a_2$. 


\bibliographystyle{apsrev4-2}
\bibliography{Bibliography}

\begin{thebibliography}{72}
\bibitem [{a_n()}]{a_noteSymmetricSpinStatesIndex}%
  \BibitemOpen
  \href@noop {} {}\bibinfo {note} {Due to the full symmetry of $\lvert 1,
  M_{F_{\mathrm{3b}}}[+] \rangle_{\alpha}$ and $\lvert 3,
  M_{F_{\mathrm{3b}}}(2) \rangle_{\alpha}$, the subscript $\alpha$ can be left
  out.}\BibitemShut {Stop}%
\end{thebibliography}

\end{document}